\documentclass[11pt, oneside]{article}
\usepackage{amsfonts}
\usepackage{mathrsfs}
\usepackage{amssymb}
\usepackage{float}
\usepackage{natbib}
\usepackage{graphicx}
\usepackage{epstopdf}
\usepackage{caption}
\usepackage{array}
\usepackage{amsmath}
\usepackage{bm}
\usepackage[shortlabels]{enumitem}
\usepackage{booktabs}
\usepackage{geometry}
\usepackage{multirow}
\usepackage{placeins}
\usepackage{latexsym,amstext}
\usepackage{mathtools}
\usepackage{amsthm}
\usepackage{indentfirst}
\usepackage{color}
\usepackage{tabularx}
\usepackage{subfigure}
\usepackage[ruled,vlined,linesnumbered]{algorithm2e}
\usepackage[colorlinks,linkcolor=blue,anchorcolor=blue,citecolor=blue]{hyperref}
\usepackage{xcolor}

\newtheorem{proposition}{Proposition}[section]

\numberwithin{equation}{section}
\allowdisplaybreaks

\newtheorem{theorem}{\color{black}\indent Theorem}[section]

\newtheorem{Assumption}{\color{black}\indent Assumption}[section]
\newtheorem{definition}{\color{black}\indent Definition}[section]

\makeatletter
\def\th@remark{%
  \normalfont 
  \let\@begintheorem\@afterindenttrue\@afterheading\flushleft 
  \let\@endtheorem\endflushleft 
}
\makeatother

\theoremstyle{plain} 
\newtheorem{remark}{\color{black}Remark}

\begin{document}

\title{Signed Matrix Thinning and Projection Estimation for Integer-Valued Autoregressive Models}
\author{Kaiyan Cui$^{1}$, Yikai Hu$^{1}$ \\
\small{$^{1}$ School of Mathematics and Statistics, Shanxi University, Taiyuan 030006, China}\\
\small{(E-mail: huyikai@sxu.edu.cn)}
}
\date{}
\maketitle

\noindent \textbf{Abstract:} Integer-valued time series are ubiquitous in fields such as finance, economics, and epidemiology. As spatiotemporal data structures in these domains grow increasingly complex and high-dimensional, the matrix integer-valued autoregressive (\textup{MINAR}) model efficiently captures row-column cross-correlations to reduce dimensionality. However, it fundamentally fails to accommodate negative values, which is a critical flaw for analyzing real-world differenced data or financial tick fluctuations. To bridge this theoretical and practical gap, this paper introduces the \textup{Z-MINAR} model, a novel matrix autoregressive framework defined on the full integer domain ($\mathbb{Z}$). By pioneering a signed matrix thinning operator and utilizing an extended poisson distribution for the innovations, the \textup{Z-MINAR} model elegantly handles both positive and negative integers while strictly preserving the crucial topological interactions inherent in matrix data. Furthermore, we employ a projection-based conditional least squares estimation procedure and rigorously establish the model's stationarity, causality, and asymptotic normality. Extensive simulations demonstrate the superior estimation accuracy, robustness, and adaptability of \textup{Z-MINAR} over existing benchmark models. Finally, an empirical application focusing on crime count variations across different urban regions confirms the model's practical efficacy in uncovering dynamic spatiotemporal dependence structures in $\mathbb{Z}$-valued matrix time series.

\vspace{0.3cm}
\noindent
\textbf{Keywords:} Negative integer time series; Relative binomial thinning operator; extended Poisson distribution; Conditional least squares estimation; Spatiotemporal dynamics

\thispagestyle{empty}
\section{Introduction}

The analysis of integer-valued time series has evolved into an indispensable branch of modern statistics, widely applied in various disciplines such as epidemiology, traffic flow monitoring, finance, and public safety management.\ As comprehensively reviewed by \citet{weiss2008thinning} and \citet{scotto2015thinning}, the fundamental challenge in modeling such data lies in the discrete nature of the state space. To circumvent this methodological hurdle, researchers have widely adopted probabilistic thinning operations. \citet{steutel1979discrete} laid the theoretical foundation for this field by defining discrete analogues of self-decomposability through the classic binomial thinning operator. Building upon this, \citet{al1987first} formally constructed the first-order integer-valued autoregressive (\textup{INAR(1)}) process. Since the introduction of this model, the \textup{INAR} series models have been extensively enriched to handle complex empirical distributions and time-varying conditional variances, with major theoretical contributions including the integer-valued generalized autoregressive conditional heteroskedasticity (INGARCH) process \citep{ferland2006integer}, Poisson autoregression \citep{fokianos2009poisson}, and various extensions of thinning mechanisms \citep{zhang2009inference, liu2020new, cao2023based, karlis2024negative}.

With the increasing complexity and dynamic nature of modern data collection technologies, the analytical focus has gradually expanded to multivariate data observed across multiple entities. Early attempts to model multidimensional integer sequences primarily relied on the multivariate generalized \textup{INAR} (\textup{MGINAR}) framework. \citet{latour1997multivariate} first established the rigorous existence criteria for a stationary and causal \textup{MGINAR} process, followed by substantial research devoted to bivariate and multivariate estimation \citep{pedeli2011bivariate, pedeli2012estimation, liu2024bivariate}. However, adapting these vector-based models to matrix observation data requires an artificial vectorization process. This transformation not only completely severs the inherent structural features and interactions between rows and columns but also leads to an exponential growth in the number of parameters to be estimated, thereby triggering a severe curse of dimensionality.

To achieve parameter parsimony while maintaining the spatiotemporal structure information of the data, researchers in the continuous data domain have advocated for the use of structurally aware matrix models \citep{walden2001wavelet, wang2017factor, chen2020autoregressive}. Inspired by these structural breakthroughs, \citet{xu2024high} innovatively constructed left- and right-multiplying matrix thinning operators, formally establishing the \textup{MINAR} model. The matrix \textup{INAR} (\textup{MINAR}) model effectively captures the cross-correlations between rows and columns and significantly reduces the number of parameters to be estimated. Recently, \citet{cui2026reduced} further expanded this framework through reduced-rank approximations to address data redundancy issues in high-dimensional environments.

However, current matrix integer-valued models face a critical limitation that severely restricts their scope of application: due to their reliance on the traditional binomial thinning operator, these models remain strictly confined to the non-negative integer domain. This non-negativity constraint creates a significant obstacle in modern empirical research. To systematically accommodate negative integers, researchers have made several theoretical breakthroughs. \citet{freeland2010true} first constructed the true \textup{INAR} (TINAR) process by viewing the process and its innovations as the difference between two independent Poisson \textup{INAR(1)} processes. This construction naturally leads to innovations that follow a Skellam distribution, which is symmetric around zero and accommodates negative counts and negative correlations. Subsequently, \citet{kachour2010p} introduced the SINAR($p$) model, which avoids the difference-of-Poisson construction by directly extending the binomial thinning operator to the entire integer domain ($\mathbb{Z}$). They proposed a ``signed thinning operator'' that multiplies the classical sum of counting sequences by the sign of the observation (i.e., $\operatorname{sign}(X) \sum_{i=1}^{|X|} Y_i$), thereby directly processing and generating negative integers while preserving the autocorrelation structure of real-valued AR processes. More recently, \citet{kachour2023new} further advanced this methodology by proposing an integer-valued process based on two novel components: a ``relative binomial thinning operator'' where the counting sequence $Y_i \in \{-1, 0, 1\}$ acts as a discrete random walk that allows steps backward into negative domain, and an extended Poisson distribution defined on $\mathbb{Z}$ for the innovations, which provides greater dispersion flexibility for modeling both positive and negative integer fluctuations.

Although the statistical community has developed these univariate and vector-based models defined on $\mathbb{Z}$, extending these models to spatiotemporal matrix scenarios presents a new challenge. As spatiotemporal data exhibit temporal trends and complex structures, researchers frequently calculate differences in data (e.g., crime counts between consecutive time periods), and these differenced datasets map onto the entire integer domain ($\mathbb{Z}$), generating positive increments and negative reductions. Directly applying vector-based models with signed thinning operators to such matrix data leads to severe parameter proliferation. Conversely, while the structurally advantageous \textup{MINAR} model efficiently captures cross-correlations to reduce dimensionality, it is completely inapplicable to differenced data with negative values. Therefore, extending the \textup{MINAR} model to handle negative values while preserving its structural advantages remains a critical yet unresolved issue.

To overcome the limitations of the aforementioned \textup{MINAR} model, this paper proposes the \textup{Z-MINAR} model, defined on the full integer domain. As the core of our methodological innovation, we generalize the theory of signed thinning to high-dimensional matrix spaces. Specifically, we introduce a signed matrix thinning operator to replace the traditional non-negative thinning operator, and integrate it with an extended Poisson distribution to precisely characterize the innovation error matrix. The proposed \textup{Z-MINAR} model effectively breaks the traditional non-negativity constraint. More importantly, by utilizing a Kronecker product structure, this model inherently preserves the cross-sectional interaction dynamics between rows and columns, achieving substantial dimension reduction while naturally accommodating negative integers. Furthermore, a projection-based conditional least squares (CLS) estimation method for the \textup{Z-MINAR} model is proposed. Specifically, the proposed projection-based CLS estimation method leverages algebraic identities rooted in singular value decomposition (SVD) \citep{denton2021eigenvectors} to project unstructured estimators onto the Kronecker manifold. Strict stationarity, ergodicity, and asymptotic normality of the resultant estimator are rigorously established under large-sample conditions. Comprehensive simulation studies illustrate that the \textup{Z-MINAR} model delivers better estimation accuracy, robustness, and adaptability relative to existing benchmark specifications. An empirical analysis based on regional crime count variation data robustly validates the effectiveness of the proposed \textup{Z-MINAR} model in uncovering the dynamic spatiotemporal dependence structures of $\mathbb{Z}$-valued matrix time series.

The remainder of this paper is organized as follows. Section 2 introduces the \textup{Z-MINAR(1)} model and establishes its fundamental probabilistic properties. Section 3 details the projection-based parameter estimation method. Section 4 presents the relevant theoretical results. Section 5 assesses the finite-sample performance of the proposed projection-based estimation via comprehensive numerical simulations, followed by an empirical analysis of regional crime data using the \textup{Z-MINAR} model. Section 6 concludes the paper. The relevant mathematical proofs and supplementary simulation results are provided in the Appendix.

\section{The \textup{Z-MINAR(1)} model}
\label{sec:z-minar1}

\subsection{Definitions}
\label{subsec:preliminary-definitions}

To establish a rigorous mathematical foundation for the matrix-variate process defined on $\mathbb{Z}$, this section first introduces two core components: (i) the extended Poisson distribution for characterizing the innovations, and (ii) the signed matrix thinning operator for constructing the autoregressive mechanism.

\subsubsection{Extended Poisson Distribution}
\label{subsubsec:extended-poisson}

Following the theoretical framework established by \citet{kachour2023new}, the innovation term of the \textup{Z-MINAR} model is assumed to follow the extended Poisson distribution, denoted as $\text{E-Po}(p, \lambda)$, where $\lambda > 0$ is the intensity parameter and $0 \leq p \leq 1$ is the probability weighting factor. Let $\epsilon\sim\text{E-Po}(p, \lambda)$ denote the integer-valued random variable following the extended Poisson distribution. Its probability mass function (PMF) is given by
\[
P(\epsilon = k) = 
\begin{cases} 
e^{-\lambda}, & k = 0, \\
pe^{-\lambda}\frac{\lambda^k}{k!}, & k = 1, 2, \ldots, \\
(1-p)e^{-\lambda}\frac{\lambda^{|k|}}{|k|!}, & k = -1, -2, \ldots.
\end{cases}
\]
The corresponding expectation, variance, and probability generating function (pgf) are given respectively by
\[
\mathbb{E}(\epsilon) = (2p-1)\lambda, \quad 
\mathbb{V}(\epsilon) = \lambda + 4p(1-p)\lambda^2, \quad 
\phi_\epsilon(s) = e^{-\lambda}\left(pe^{s\lambda} + (1-p)e^{\lambda/s}\right).
\]
The extended Poisson distribution can accommodate both positive and negative integer-valued innovations, which facilitates the construction of time series models suited for real-world differenced data.

\subsubsection{Signed Matrix Thinning Operator}
\label{subsubsec:signed-matrix-thinning}

By combining the concept of signed thinning with the structured matrix thinning architecture \citep{xu2024high,cui2026reduced}, we formally define the signed matrix thinning operator $\circledast_s$ in Definition \ref{def:signed-matrix-thinning}.

\begin{definition}\label{def:signed-matrix-thinning}
Let $\mathbf{A} = (\alpha_{i,j}) \in \mathbb{R}_{\geq0}^{m \times m}$ and $\mathbf{B} = (\beta_{k,l}) \in \mathbb{R}_{\geq0}^{n \times n}$ be non-negative matrices, and let $\mathbf{Y} = (y_{i,j}) \in \mathbb{Z}^{m \times n}$ be a $\mathbb{Z}$-valued random matrix. By defining the standard sign function $\operatorname{sign}(\cdot)$ $(\operatorname{sign}(x) = 1$ if $x > 0$, $\operatorname{sign}(x) = -1$ if $x < 0$, and $\operatorname{sign}(0) = 0)$ and the conventional binomial thinning operator $\alpha \circ y$ {\rm(}for a non-negative integer $y$, $\alpha \circ y = \sum_{i=1}^{y} \xi_i^{(\alpha)}$, where $\xi_i^{(\alpha)}$ is a sequence of independent and identically distributed Bernoulli random variables with $P(\xi_i = 1) = \alpha$ and $P(\xi_i = 0) = 1-\alpha${\rm)}, and assuming the thinning operations are mutually independent across all row indices $i, k$ and column indices $j, l$, the signed matrix thinning operator is defined as:
\[
\mathbf{A} \circledast_s \mathbf{Y} \circledast_s \mathbf{B}^\top = \left( \sum_{l=1}^n \sum_{k=1}^m \operatorname{sign}(y_{k,l}) \cdot (\beta_{j,l}\alpha_{i,k}) \circ |y_{k,l}| \right)_{1 \leq i \leq m, 1 \leq j \leq n}.
\]
\end{definition}

To explicitly formalize the corresponding vector-level operations, we define the signed vector thinning operator $\ast_s$:

\begin{definition}
For a $P \times P$ dimensional matrix $\mathbf{M} = (\mu_{i,j})_{P \times P} \in \mathbb{R}_{\geq0}^{P \times P}$ and a $\mathbb{Z}$-valued random vector $\mathbf{v} = (v_1, \ldots, v_P)^\top \in \mathbb{Z}^P$, the signed vector thinning operator $\ast_s$ is defined as
\[
\mathbf{M} \ast_s \mathbf{v} := \left( \sum_{j=1}^P \operatorname{sign}(v_j) \cdot (\mu_{i,j} \circ |v_j|) \right)_{1 \leq i \leq P},
\]
where for $1 \leq i, j \leq P$, the thinning operators $\mu_{i,j} \circ \cdot$ operate independently.
\end{definition}

\begin{proposition}
\label{property:vec-signed-thinning}
The signed matrix thinning operator $\circledast_s$ satisfies the vectorization identity
$\operatorname{vec}(\mathbf{A} \circledast_s \mathbf{Y} \circledast_s \mathbf{B}^\top) = (\mathbf{B} \otimes \mathbf{A}) \ast_s \operatorname{vec}(\mathbf{Y})$, where $\otimes$ denotes the Kronecker product.
\end{proposition}

\subsection{Model}
\label{subsec:model-definition}

Based on Definition \ref{def:signed-matrix-thinning}, we develop the framework of the \textup{Z-MINAR(1)} model, which is formally introduced in Definition \ref{def:zminar}.

\begin{definition}\label{def:zminar}
Let $\mathbf{A} \in \mathbb{R}_{\geq0}^{m \times m}$ and $\mathbf{B} \in \mathbb{R}_{\geq0}^{n \times n}$ be the row and column interaction coefficient matrices, respectively. Let $\mathbf{E}_t = (e_{t, i,j}) \in \mathbb{Z}^{m \times n}$ be the innovation matrix, whose elements are mutually independent and follow the extended Poisson distribution $\text{E-Po}(p_{i,j}, \lambda_{i,j})$. Assume that for any $t$, the historical realization $\mathbf{X}_{t-1}$ and the concurrent innovation $\mathbf{E}_t$ are mutually independent, and the thinning operator $\circledast_s$ is executed independently across different time steps $t$. A sequence of $m \times n$-dimensional $\mathbb{Z}$-valued random matrices $\{\mathbf{X}_t\}_{t \in \mathbb{Z}}$ is said to follow a \textup{Z-MINAR(1)} process if it satisfies the following stochastic difference equation:
\[
\mathbf{X}_t = \mathbf{A} \circledast_s \mathbf{X}_{t-1} \circledast_s \mathbf{B}^\top + \mathbf{E}_t, \quad t \in \mathbb{Z}.
\]
\end{definition}

By Proposition \ref{property:vec-signed-thinning}, there exists an interchangeable relationship between the \textup{Z-MINAR(1)} and \textup{MGINAR(1)} models:
\begin{equation}
\operatorname{vec}(\mathbf{X}_t) = \mathbf{\Phi} \ast_s \operatorname{vec}(\mathbf{X}_{t-1}) + \operatorname{vec}(\mathbf{E}_t),
\end{equation}
where $\mathbf{\Phi} = \mathbf{B} \otimes \mathbf{A}$.

\subsection{Probabilistic Properties}
\label{subsec:probabilistic-properties}

In this section, we state the existence, strict stationarity and ergodicity of the proposed \textup{Z-MINAR(1)} model.

\begin{proposition}[Stationarity and Causality Criterion]
\label{prop:stationarity-causality}
If the product of the spectral radii of the coefficient matrices $\mathbf{A}$ and $\mathbf{B}$ satisfies $\rho(\mathbf{A})\rho(\mathbf{B}) < 1$, the generated \textup{Z-MINAR(1)} process is stationary and causal. Furthermore, to ensure the probabilistic validity of the binomial thinning operations in Definition \ref{def:signed-matrix-thinning}, it is required that all elements of the Kronecker product matrix $\mathbf{\Phi} = \mathbf{B} \otimes \mathbf{A}$ fall within the closed interval $[0, 1]$ {\rm(}i.e., $\max_{i,k}\alpha_{i,k}\max_{j,l}\beta_{j,l} \leq 1${\rm)}.
\end{proposition}

Under the conditions of Proposition \ref{prop:stationarity-causality}, derivations for the conditional expectation, unconditional expectation and variance-covariance matrix closely follow the framework developed by \citet{xu2024high}. Our \textup{Z-MINAR(1)} model differs primarily through the signed thinning operator and generalized Poisson innovations. These components modify $\mathbf{C} = \mathbb{E}(\mathbf{E}_t)$ and the variance structure of the innovation matrix, whereas the macroscopic matrix formulation and interaction dynamics are preserved. Accordingly, we omit redundant proofs and summarize the probabilistic properties of the \textup{Z-MINAR(1)} model below:
\begin{itemize}
    \item \textbf{Conditional Expectation and Unconditional Expectation:} By Definitions \ref{def:signed-matrix-thinning} and \ref{def:zminar}, we have
    \[ \mathbb{E}(\mathbf{X}_t|\mathbf{X}_{t-1}) = \mathbf{A} \mathbf{X}_{t-1} \mathbf{B}^\top + \mathbf{C}, \quad t \in \mathbb{Z}. \]
    Let $\bm{\mu}_X := \mathbb{E}(\mathbf{X}_t)$ be the unconditional expectation matrix of $\mathbf{X}_t$. Then its vectorized form is given by:
    \[ \operatorname{vec}(\bm{\mu}_X) = \left(\mathbf{I}_{mn \times mn} - \mathbf{B} \otimes \mathbf{A}\right)^{-1} \operatorname{vec}(\mathbf{C}), \quad t \in \mathbb{Z}, \]
    where the innovation expectation matrix is $\mathbf{C} = \mathbb{E}(\mathbf{E}_t)$ with elements $c_{i,j} = (2p_{i,j}-1)\lambda_{i,j}$. This explicitly allows for negative drift, distinguishing it from the strictly positive framework in \citet{xu2024high}.
    
    \item \textbf{Variance-Covariance Matrix:}
    Let $\mathbf{\Gamma}_h := \operatorname{Cov}(\operatorname{vec}(\mathbf{X}_t), \operatorname{vec}(\mathbf{X}_{t-h}))$ denote the autocovariance matrix at lag-$h$. Write $\mathcal{T} = (\tau_{i,j})_{mn \times mn}$, where $\tau_{i,j} = 1$ if $i \in \{sm+1, sm+2, \dots, (s+1)m\}$, $j \in \{(s+1) + (i-sm-1)n\}$ and $s \in \{0, 1, \dots, n-1\}$, and $\tau_{i,j} = 0$ otherwise. Similiar to \citet{xu2024high}, the column-wise and row-wise variance-covariance matrices of $\mathbf{X}_t$ are respectively given by
    \[ \mathbf{\Sigma}_0^c = \mathcal{T}\mathbf{\Gamma}_0^{\otimes}, \quad \mathbf{\Sigma}_0^r = \mathbf{\Gamma}_0^{\otimes}\mathcal{T}, \]
    where $\mathbf{\Gamma}_0^{\otimes} = \mathbb{E}\left((\mathbf{X}_t - \bm{\mu}_X) \otimes (\mathbf{X}_t - \bm{\mu}_X)^\top\right)$. The distinctive feature in the \textup{Z-MINAR(1)} model is that the innovation variance contribution is governed by the extended Poisson distribution, $\operatorname{Var}(e_{t, i,j}) = \lambda_{i,j} + 4p_{i,j}(1-p_{i,j})\lambda_{i,j}^2$, providing a more flexible dispersion structure compared to standard non-negative count models. The column-wise and row-wise autocovariance matrices with lag-$h$ are $\mathbf{\Sigma}_h^c = \mathcal{T}\mathbf{\Gamma}_h^{\otimes}$ and  $\mathbf{\Sigma}_h^r = \mathbf{\Gamma}_h^{\otimes}\mathcal{T}$, respectively.
\end{itemize}

\subsection{Model Interpretation}

To intuitively interpret the practical motivation and interaction mechanisms of the \textup{Z-MINAR} model, we examine the genesis of \(\mathbb{Z}\)-valued data and the intrinsic structure of matrix observations. In many empirical scenarios, such as spatiotemporal crime analysis and traffic monitoring, raw data take the form of non-negative matrix counts \(\mathbf{Y}_t\). In practice, such non-negative integer matrix series commonly exhibit non-stationary behaviors with evident temporal variations. As extensively discussed in \citet{kachour2023new}, \(\mathbb{Z}\)-valued fluctuations are prevalent in real-world systems, including intraday stock price movements in financial markets, goal differentials in sports matches, and changes in disease symptom counts in medical studies.

We next introduce the temporal differencing operator \(\nabla\), defined as \(\nabla \mathbf{Y}_t = \mathbf{Y}_t - \mathbf{Y}_{t-1}\), and define the differenced series as \(\mathbf{X}_t = \nabla \mathbf{Y}_t\). This transformation maps non-negative count data onto the entire integer domain \(\mathbb{Z}\). For non-stationary non-negative integer time series, differencing serves as a standard tool for stationarization, naturally yielding \(\mathbb{Z}\)-valued processes. This differencing mechanism is mathematically analogous to the integrated component ``I'' in the classical ARIMA framework. Therefore, if the $d$-th order differenced matrix series $\nabla^d \mathbf{Y}_t$ follows a \textup{Z-MINAR}($p$) process, we formally define the original non-negative matrix series \(\mathbf{Y}_t\) as following an integrated \textup{Z-MIINAR}($p, d$) model. This explicit notation standardizes the theoretical framework and highlights the model's practical significance, enabling the characterization of non-stationary growth, reduction, and dynamic volatility of matrix count data across diverse research fields.

For such matrix-structured differenced data, complex cross-correlations naturally exist across both rows and columns. Take spatiotemporal crime data for illustration: suppose rows correspond to distinct crime categories (e.g., battery, other offense, and theft), and columns denote separate geographic districts. Row-wise connections emerge due to underlying socioeconomic or environmental factors. For instance, an increase in battery offenses in an area might be strongly correlated with a simultaneous rise in thefts, reflecting a general deterioration of local security. Functionally, the left-multiplying matrix $\mathbf{A}$ captures these dynamic interactions between rows (crime types). Meanwhile, column-wise connections arise from spatial spillover effects or geographic proximity. For example, an outbreak of crimes in one district can easily spill over into neighboring districts, as offenders cross boundaries or policing strategies shift. The right-multiplying matrix $\mathbf{B}$ governs these cross-sectional interactions between columns (spatial districts).

Crucially, in complex spatiotemporal systems, these row-wise typological correlations and column-wise spatial spillovers do not operate in isolation; rather, they are deeply intertwined and evolve concurrently. Imposing traditional vector-based frameworks (such as \textup{MGINAR} or \textup{Z-VINAR}) on such inherently two-dimensional data necessitates an artificial vectorization procedure. This transformation not only obliterates the underlying topological structure but also precipitates a severe parameter proliferation, thereby exacerbating the curse of dimensionality. Consequently, preserving the matrix-variate representation is both mathematically rigorous and empirically indispensable. By formulating the model with bilateral coefficient matrices---where the left-multiplying matrix encapsulates row-wise dynamics and the right-multiplying matrix governs column-wise interactions, we can simultaneously delineate dual-dimensional dependencies without sacrificing structural integrity. This parsimonious parameterization effectively averts structural information loss and achieves substantial dimension reduction, thereby fundamentally motivating the architectural design of the proposed \textup{Z-MINAR} framework.

\begin{remark}
Under the generalized \textup{Z-MINAR} framework, the operation $\mathbf{A} \circledast_s \mathbf{X}_{t-1} \circledast_s \mathbf{B}^{\top}$ elegantly captures these composite row-column interaction effects inherent in the differenced data. The intrinsic Kronecker product structure $\mathbf{B} \otimes \mathbf{A}$ significantly compresses the parameter space, achieving substantial dimension reduction while perfectly maintaining the multidimensional structural integrity of the integrated time series.
\end{remark}

\subsection{Vectorized Representation}
\label{subsec:vectorization}

To facilitate parameter estimation and theoretical derivation, we first formulate the \textup{Z-MINAR(1)} model as a structured vector autoregressive model. This is achieved through the following white noise representation. 

\begin{proposition}
\label{prop:white-noise}
Let $\{\mathbf{X}_t\}_{t\in\mathbb{Z}}$ be a matrix sequence generated from an $m \times n$-dimensional \textup{Z-MINAR(1)} model with coefficient matrices $\mathbf{A} \in \mathbb{R}_{\geq 0}^{m \times m}$ and $\mathbf{B} \in \mathbb{R}_{\geq 0}^{n \times n}$ such that $\rho(\mathbf{A})\rho(\mathbf{B}) < 1$ holds and innovation-parameter matrix $\mathbf{C} \in \mathbb{R}_{\geq 0}^{m \times n} \backslash \{\mathbf{0}_{m \times n}\}$. Then
\[
\mathbf{U}_t := \mathbf{X}_t - \mathbf{A}\mathbf{X}_{t-1}\mathbf{B}^\top - \mathbf{C}, \quad t \in \mathbb{Z},
\]
defines a matrix white noise sequence with concurrent correlations among its own entries, i.e., $\{\mathbf{U}_t\}_{t\in\mathbb{Z}}$ is stationary, $\mathbb{E}\mathbf{U}_t = \mathbf{0}_{m \times n}$, $t \in \mathbb{Z}$, and
\[
\mathbb{E}\left( \operatorname{vec}(\mathbf{U}_{t_1}) \operatorname{vec}(\mathbf{U}_{t_2})^\top \right) = 
\begin{cases}
\mathbf{\Sigma}_U, & t_1 = t_2 \\
\mathbf{0}_{mn \times mn}, & t_1 \neq t_2
\end{cases}
\]
where $\mathbf{\Sigma}_U = \operatorname{diag}\left( [(\mathbf{B} \otimes \mathbf{A}) \circ (\mathbf{J}_{mn \times mn} - \mathbf{B} \otimes \mathbf{A})] \mathbb{E}(|\operatorname{vec}(\mathbf{X}_{t-1})|) \right) + \mathbf{\Sigma}_E$, with $\mathbf{J}_{mn \times mn}$ being an $mn \times mn$ matrix of ones and $\mathbf{\Sigma}_E = \operatorname{Cov}(\operatorname{vec}(\mathbf{E}_t))$ being the variance of the innovation.
\end{proposition}

Based on Proposition \ref{prop:white-noise}, the model can be rewritten as:
\[
\mathbf{X}_t = \mathbf{A} \mathbf{X}_{t-1} \mathbf{B}^\top + \mathbf{C} + \mathbf{U}_t, \quad t \in \mathbb{Z}.
\]
This representation effectively replaces the signed matrix thinning operator $\circledast_s$ in the original model with conventional matrix multiplication. Let $\mathbf{\Phi} = \mathbf{B} \otimes \mathbf{A}$. Vectorizing the equation yields the standard VAR(1) form:
\[
\operatorname{vec}(\mathbf{X}_t) = \mathbf{\Phi} \operatorname{vec}(\mathbf{X}_{t-1}) + \operatorname{vec}(\mathbf{C}) + \operatorname{vec}(\mathbf{U}_t).
\]

\subsection{Parameter Identifiability}
\label{subsec:identifiability}
A critical feature of the proposed \textup{Z-MINAR} framework is the structural parameterization via the Kronecker product $\mathbf{\Phi} = \mathbf{B} \otimes \mathbf{A}$. However, this bilinear specification introduces an inherent scale ambiguity. For any arbitrary non-zero scalar $c$, the transformation $\tilde{\mathbf{A}} = c\mathbf{A}$ and $\tilde{\mathbf{B}} = \frac{1}{c}\mathbf{B}$ yields an identical Kronecker product $\tilde{\mathbf{B}} \otimes \tilde{\mathbf{A}} = (\frac{1}{c}\mathbf{B}) \otimes (c\mathbf{A}) = \mathbf{B} \otimes \mathbf{A}$. This shifting of scale leads to an infinite number of parameter combinations that yield the exact same conditional expectation, rendering the unconstrained system non-identifiable.

To ensure unique identifiability of the model parameters and eliminate estimation ambiguity---\allowbreak akin to the structural identifiability issues discussed in recent matrix integer-valued frameworks \citep{cui2026additive}---\allowbreak we impose the strict scale normalization constraint $\|\mathbf{A}\|_F = 1$. Under this constraint, the coefficient matrices $\mathbf{A}$ and $\mathbf{B}$ can be uniquely determined from the global autoregressive matrix $\mathbf{\Phi}$.

\section{Parameter Estimation}
\label{sec:estimation}

In this section, we propose a projection-based conditional least squares estimation (PROJ) methodology for the parameter matrices $\mathbf{A}$, $\mathbf{B}$, and $\mathbf{C}$ within the \textup{Z-MINAR(1)} model. The procedure is divided into two sequential steps: first obtaining an unconstrained global estimator, and subsequently projecting it onto the structurally constrained parameter space to derive explicit structural estimates.

\subsection{Conditional Least Squares Estimation}
Based on the vectorized representation in Section \ref{subsec:vectorization}, the conditional expectation of the model is given by:
\[
\mathbb{E}(\operatorname{vec}(\mathbf{X}_t)\mid\operatorname{vec}(\mathbf{X}_{t-1})) = \mathbf{\Phi}\operatorname{vec}(\mathbf{X}_{t-1}) + \operatorname{vec}(\mathbf{C}) = \mathbf{\Psi}\mathbf{z}_{t-1},
\]
where the augmented parameter matrix is $\mathbf{\Psi} = [\mathbf{\Phi} \vdots \operatorname{vec}(\mathbf{C})] \in \mathbb{R}^{mn \times (mn+1)}$, and the predictor vector is $\mathbf{z}_{t-1} = (\operatorname{vec}(\mathbf{X}_{t-1})^\top, 1)^\top \in \mathbb{R}^{(mn+1) \times 1}$. 

Given a sample of length $T$, we define the conditional sum of squared residuals as the objective function $Q_T(\mathbf{\Psi})$:
\[
Q_T(\mathbf{\Psi}) = \sum_{t=2}^T \left\| \operatorname{vec}(\mathbf{X}_t) - \mathbf{\Psi} \mathbf{z}_{t-1} \right\|_F^2 = \sum_{t=2}^T \operatorname{tr}\left( (\operatorname{vec}(\mathbf{X}_t) - \mathbf{\Psi} \mathbf{z}_{t-1})(\operatorname{vec}(\mathbf{X}_t) - \mathbf{\Psi} \mathbf{z}_{t-1})^\top \right).
\]
To obtain the unconstrained conditional least squares estimate $\hat{\mathbf{\Psi}}_{\mathit{LS}}$ without the Kronecker structure constraint, we minimize $Q_T(\mathbf{\Psi})$. Taking the partial derivative of $Q_T(\mathbf{\Psi})$ with respect to the matrix $\mathbf{\Psi}$ and setting it to zero yields:
\[
\frac{\partial Q_T(\mathbf{\Psi})}{\partial \mathbf{\Psi}} = -2 \sum_{t=2}^T (\operatorname{vec}(\mathbf{X}_t) - \mathbf{\Psi} \mathbf{z}_{t-1}) \mathbf{z}_{t-1}^\top = \mathbf{0}.
\]
Rearranging the above equation, we obtain the normal equations:
\[
\hat{\mathbf{\Psi}}_{\mathit{LS}} \left( \sum_{t=2}^T \mathbf{z}_{t-1} \mathbf{z}_{t-1}^\top \right) = \sum_{t=2}^T \operatorname{vec}(\mathbf{X}_t) \mathbf{z}_{t-1}^\top.
\]
Post-multiplying both sides by the inverse of the sample covariance matrix of the predictors, we get the closed-form solution:
\[
\hat{\mathbf{\Psi}}_{\mathit{LS}} = \left( \sum_{t=2}^T \operatorname{vec}(\mathbf{X}_t) \mathbf{z}_{t-1}^\top \right) \left( \sum_{t=2}^T \mathbf{z}_{t-1} \mathbf{z}_{t-1}^\top \right)^{-1}.
\]
The unconstrained global estimates $\hat{\mathbf{\Phi}}_{\mathit{LS}}$ and $\operatorname{vec}(\hat{\mathbf{C}}_{\mathit{LS}})$ correspond to the first $mn$ columns and the final column of $\hat{\mathbf{\Psi}}_{\mathit{LS}}$, respectively. Denoting $\mathcal{R}_{m,n}$ as the reshaping operator that maps a vectorized input back into its corresponding $m \times n$ matrix format, the intercept matrix estimator is given by:
\[
\hat{\mathbf{C}}_{\mathit{LS}} = \mathcal{R}_{m,n}(\hat{\mathbf{\Psi}}_{\mathit{LS}, 1:mn,\, mn+1}).
\]

\subsection{Projection Estimation Method}
The unconstrained matrix $\hat{\mathbf{\Phi}}_{\mathit{LS}}$ does not inherently satisfy the Kronecker product structure $\mathbf{B} \otimes \mathbf{A}$. We extract the structural estimators by projecting $\hat{\mathbf{\Phi}}_{\mathit{LS}}$ onto the Kronecker manifold under the Frobenius norm:
\begin{align*}
(\hat{\mathbf{A}}_{\mathit{PROJ}}, \hat{\mathbf{B}}_{\mathit{PROJ}}) &= \arg\min_{\mathbf{A},\mathbf{B}} \| \hat{\mathbf{\Phi}}_{\mathit{LS}} - \mathbf{B} \otimes \mathbf{A} \|_F^2, \\
\hat{\mathbf{C}}_{\mathit{PROJ}} &= \hat{\mathbf{C}}_{\mathit{LS}} = \mathcal{R}_{m,n}(\hat{\mathbf{\Psi}}_{\mathit{LS}, 1:mn,\, mn+1}).
\end{align*}
This minimization is mathematically equivalent to the nearest Kronecker product (NKP) problem \citep{van2000ubiquitous}. To solve it, we introduce the rearrangement operator $\mathcal{G}: \mathbb{R}^{mn \times mn} \rightarrow \mathbb{R}^{m^2 \times n^2}$ defined such that $\mathcal{G}(\mathbf{B} \otimes \mathbf{A}) = \operatorname{vec}(\mathbf{A}) \operatorname{vec}(\mathbf{B})^\top$. Applying this operator to $\hat{\mathbf{\Phi}}_{\mathit{LS}}$ yields $\tilde{\mathbf{\Phi}} = \mathcal{G}(\hat{\mathbf{\Phi}}_{\mathit{LS}})$.

Since the Frobenius norm is defined as the square root of the sum of the absolute squares of its elements, it is invariant to the spatial arrangement of the elements. Therefore, we can rewrite the objective function as:
\begin{align*}
\| \hat{\mathbf{\Phi}}_{\mathit{LS}} - \mathbf{B} \otimes \mathbf{A} \|_F^2 &= \sum \left( \hat{\mathbf{\Phi}}_{\mathit{LS}} - \mathbf{B} \otimes \mathbf{A} \right)^2 \\
&= \sum \left( \tilde{\mathbf{\Phi}} - \operatorname{vec}(\mathbf{A}) \operatorname{vec}(\mathbf{B})^\top \right)^2 \\
&= \| \tilde{\mathbf{\Phi}} - \operatorname{vec}(\mathbf{A}) \operatorname{vec}(\mathbf{B})^\top \|_F^2.
\end{align*}
Thus, the problem transforms into finding the best rank-1 approximation of the rearranged matrix $\tilde{\mathbf{\Phi}}$:
\[
\min_{\mathbf{A},\mathbf{B}} \| \tilde{\mathbf{\Phi}} - \operatorname{vec}(\mathbf{A}) \operatorname{vec}(\mathbf{B})^\top \|_F^2.
\]
According to the Eckart-Young-Mirsky theorem, the optimal rank-1 approximation is obtained via the leading term of the singular value decomposition (SVD). Let the SVD of $\tilde{\mathbf{\Phi}}$ be $\tilde{\mathbf{\Phi}} = \sum_{k} d_k \mathbf{u}_k \mathbf{v}_k^\top$, where $d_1 \geq d_2 \geq \dots \geq 0$ are the singular values, and $\mathbf{u}_k \in \mathbb{R}^{m^2}$, $\mathbf{v}_k \in \mathbb{R}^{n^2}$ are the corresponding left and right singular vectors. The optimal solution must satisfy:
\[
\operatorname{vec}(\hat{\mathbf{A}}_{\mathit{PROJ}}) \operatorname{vec}(\hat{\mathbf{B}}_{\mathit{PROJ}})^\top = d_1 \mathbf{u}_1 \mathbf{v}_1^\top.
\]
Incorporating the identifiability constraint $\|\mathbf{A}\|_F = 1$, and since $\|\mathbf{u}_1\|_2 = 1$ is naturally guaranteed by standard SVD, we can explicitly define the vectorized projection estimators as $\operatorname{vec}(\hat{\mathbf{A}}_{\mathit{PROJ}}) = \mathbf{u}_1$ and $\operatorname{vec}(\hat{\mathbf{B}}_{\mathit{PROJ}}) = d_1 \mathbf{v}_1$.

Finally, by reshaping the vectors back into matrices and truncating any potential negative values (denoted by the element-wise operation $x_+ = \max(x, 0)$), we obtain the explicit formulas for the PROJ estimators:
\begin{align*}
\hat{\mathbf{A}}_{\mathit{PROJ}} &= \mathcal{R}_{m,m}(\mathbf{u}_1)_+, \\
\hat{\mathbf{B}}_{\mathit{PROJ}} &= \mathcal{R}_{n,n}(d_1 \mathbf{v}_1)_+, \\
\hat{\mathbf{C}}_{\mathit{PROJ}} &= \hat{\mathbf{C}}_{\mathit{LS}} = \mathcal{R}_{m,n}(\hat{\mathbf{\Psi}}_{\mathit{LS}, 1:mn,\, mn+1}).
\end{align*}
This explicit boundary projection is a common and necessary regularization practice when estimating non-negative parameters via unconstrained conditional least squares methods. It ensures the empirical validity of the spatial and cross-sectional interactions ($\mathbb{R}_{\geq 0}^{m \times m}$ and $\mathbb{R}_{\geq 0}^{n \times n}$) without invalidating the overall asymptotic framework. Meanwhile, the intercept matrix $\hat{\mathbf{C}}_{\mathit{PROJ}}$ intrinsically remains unconstrained to comfortably accommodate both positive and negative drifts. Note that under the assumption $a_{1,1} > 0$ and $\mathbf{B} \neq \mathbf{0}$, if $\mathbf{B}$ or $\mathbf{A}$ contains any negative values, their Kronecker product $\mathbf{\Phi} = \mathbf{B} \otimes \mathbf{A}$ would inevitably contain negative values. Thus, the non-negativity of $\mathbf{A}$ and $\mathbf{B}$ is mathematically equivalent to the non-negativity of $\mathbf{\Phi}$, implying that this explicit truncation procedure on the individual parameter matrices is equivalent to imposing a non-negativity constraint on the global estimator.

We illustrate the re-arrangement operation with a special case of $m=n=2$. We first rearrange the entries of the Kronecker product $\mathbf{B} \otimes \mathbf{A}$:
\[
\begin{bmatrix}
b_{1,1}a_{1,1} & b_{1,1}a_{1,2} & b_{1,2}a_{1,1} & b_{1,2}a_{1,2} \\
b_{1,1}a_{2,1} & b_{1,1}a_{2,2} & b_{1,2}a_{2,1} & b_{1,2}a_{2,2} \\
b_{2,1}a_{1,1} & b_{2,1}a_{1,2} & b_{2,2}a_{1,1} & b_{2,2}a_{1,2} \\
b_{2,1}a_{2,1} & b_{2,1}a_{2,2} & b_{2,2}a_{2,1} & b_{2,2}a_{2,2}
\end{bmatrix} \longrightarrow
\begin{bmatrix}
b_{1,1}a_{1,1} & b_{2,1}a_{1,1} & b_{1,2}a_{1,1} & b_{2,2}a_{1,1} \\
b_{1,1}a_{2,1} & b_{2,1}a_{2,1} & b_{1,2}a_{2,1} & b_{2,2}a_{2,1} \\
b_{1,1}a_{1,2} & b_{2,1}a_{1,2} & b_{1,2}a_{1,2} & b_{2,2}a_{1,2} \\
b_{1,1}a_{2,2} & b_{2,1}a_{2,2} & b_{1,2}a_{2,2} & b_{2,2}a_{2,2}
\end{bmatrix}.
\]
We then rearrange the entries of $\hat{\mathbf{\Phi}}_{\mathit{LS}}$ in exactly the same way:
\[
\hat{\mathbf{\Phi}}_{\mathit{LS}} = \begin{bmatrix}
\phi_{1,1} & \phi_{1,2} & \phi_{1,3} & \phi_{1,4} \\
\phi_{2,1} & \phi_{2,2} & \phi_{2,3} & \phi_{2,4} \\
\phi_{3,1} & \phi_{3,2} & \phi_{3,3} & \phi_{3,4} \\
\phi_{4,1} & \phi_{4,2} & \phi_{4,3} & \phi_{4,4}
\end{bmatrix} \longrightarrow
\begin{bmatrix}
\phi_{1,1} & \phi_{3,1} & \phi_{1,3} & \phi_{3,3} \\
\phi_{2,1} & \phi_{4,1} & \phi_{2,3} & \phi_{4,3} \\
\phi_{1,2} & \phi_{3,2} & \phi_{1,4} & \phi_{3,4} \\
\phi_{2,2} & \phi_{4,2} & \phi_{2,4} & \phi_{4,4}
\end{bmatrix} =: \tilde{\mathbf{\Phi}}.
\]

\section{Theory}
\label{sec:theory}

This section systematically establishes the core theoretical properties of the \textup{Z-MINAR(1)} model. Drawing upon the rigorous existence criteria for multivariate integer-valued processes pioneered by \citet{latour1997multivariate}, and considering the theoretical bounds of asymptotic behavior in complex count trajectories \citep{barczy2010asymptotic}, we first formulate the white noise representation of the model. To ensure rigorous statistical inference, we also explicitly outline the necessary regularity conditions and establish the strong consistency of the estimators before proving their asymptotic normality.

\subsection{Regularity Conditions}
\label{subsec:regularity}

\begin{Assumption}[Parameter Space]
\label{assum:parameter_space}
The true parameter matrices $\mathbf{A} \in \mathbb{R}^{m \times m}$, $\mathbf{B} \in \mathbb{R}^{n \times n}$, and the constant drift matrix $\mathbf{C} \in \mathbb{R}^{m \times n}$ lie strictly within the interior of a compact parameter space $\mathbf{\Theta}$. Furthermore, to avoid boundary complications during asymptotic derivation, we assume $\mathbf{A} \in \mathbb{R}_{>0}^{m \times m}$ and $\mathbf{B} \in \mathbb{R}_{>0}^{n \times n}$.
\end{Assumption}

\begin{Assumption}[Stationarity and Ergodicity]
\label{assum:stationarity}
The product of the spectral radii satisfies $\rho(\mathbf{A})\rho(\mathbf{B}) < 1$. This condition strictly guarantees that the generated matrix time series $\{\mathbf{X}_t\}_{t \in \mathbb{Z}}$ is stationary and ergodic.
\end{Assumption}

\begin{Assumption}[Moment Condition]
\label{assum:moment}
The extended Poisson innovation matrix $\mathbf{E}_t$ possesses finite second moments, satisfying $\mathbb{E}\|\mathbf{E}_t\|_F^2 < \infty$. This is a fundamental prerequisite for the application of conditional least squares estimation.
\end{Assumption}

\begin{Assumption}[Identifiability]
\label{assum:identifiability}
To uniquely identify the structural parameter matrices from the Kronecker product manifold $\mathbf{B} \otimes \mathbf{A}$ and eliminate intrinsic scale ambiguity, we enforce the scale normalization constraint $\|\mathbf{A}\|_F = 1$.
\end{Assumption}

\subsection{White Noise Representation}
\label{subsec:white_noise}

\begin{theorem}[White Noise Representation]\label{thm:white_noise}
Let $\{\mathbf{X}_t\}_{t\in\mathbb{Z}}$ be a stationary matrix sequence generated by the \textup{Z-MINAR(1)} model, and define the residual matrix as:
\[
\mathbf{U}_t = \mathbf{X}_t - \mathbf{A} \mathbf{X}_{t-1} \mathbf{B}^\top - \mathbf{C},\quad t\in\mathbb{Z}.
\]
Then, $\{\mathbf{U}_t\}_{t\in\mathbb{Z}}$ constitutes a stationary matrix white noise sequence that satisfies the following properties:
\begin{itemize}
    \item[(i)] $\mathbb{E}(\mathbf{U}_t) = \mathbf{0}_{m\times n}$;
    \item[(ii)] For any $t \neq s$, the temporal cross-covariance $\operatorname{Cov}(\operatorname{vec}(\mathbf{U}_t), \operatorname{vec}(\mathbf{U}_s)) = \mathbf{0}$;
    \item[(iii)] Let $\mathbf{\Sigma}_E = \operatorname{diag}\big(\sigma_{1}^2, \ldots, \sigma_{mn}^2\big)$ be the innovation variance with $\sigma_{k}^2 = \lambda_{k} + 4p_{k}(1-p_{k})\lambda_{k}^2$, where $p_k$ and $\lambda_k$ denote the corresponding parameters of the $k$-th element of the vectorized innovation $\operatorname{vec}(\mathbf{E}_t)$. Let $\mathbf{x}_{t-1} = \operatorname{vec}(\mathbf{X}_{t-1})$ be the vectorized lagged state, and let $\mathbf{\Phi} = \mathbf{B} \otimes \mathbf{A}$ be the $mn \times mn$ autoregressive coefficient matrix. Let $\mathbf{\Phi}^{\circ 2}$ denote the Hadamard product (element-wise product) of $\mathbf{\Phi}$ with itself, such that $\mathbf{\Phi} \circ (\mathbf{J} - \mathbf{\Phi})$ has elements $\phi_{i,j}(1-\phi_{i,j})$, where $\mathbf{J}$ is the $mn \times mn$ matrix of ones. The thinning variance matrix $\mathbf{\Sigma}_{thin}$ is a diagonal matrix defined as:
    \[
    \mathbf{\Sigma}_{thin} = \operatorname{diag}\left( [\mathbf{\Phi} \circ (\mathbf{J} - \mathbf{\Phi})] \mathbb{E}(|\mathbf{x}_{t-1}|) \right),
    \]
    where $|\mathbf{x}_{t-1}|$ denotes the element-wise absolute value of the vector $\mathbf{x}_{t-1}$.
    The unconditional covariance $\operatorname{Cov}(\operatorname{vec}(\mathbf{U}_t)) = \mathbf{\Sigma}_U$ is a strictly diagonal matrix given by:
    \[
    \mathbf{\Sigma}_U = \mathbf{\Sigma}_{thin} + \mathbf{\Sigma}_E.
    \]
\end{itemize}
Additionally, for any $t_1 < t$, $\operatorname{Cov}(\operatorname{vec}(\mathbf{U}_t), \operatorname{vec}(\mathbf{X}_{t_1})) = \mathbf{0}$.
\end{theorem}

\subsection{Asymptotic Properties of the Estimators}
\label{subsec:asymptotic}

To establish the asymptotic properties of the PROJ estimators $\hat{\mathbf{A}}_{\mathit{PROJ}}$, $\hat{\mathbf{B}}_{\mathit{PROJ}}$, and $\hat{\mathbf{C}}_{\mathit{PROJ}}$, we first present the asymptotic distribution of the unconstrained conditional least squares estimators for the fully vectorized model. Let $\mathbf{\Psi} = [\mathbf{\Phi}, \operatorname{vec}(\mathbf{C})]$ where $\mathbf{\Phi} = \mathbf{B} \otimes \mathbf{A}$. The unconstrained estimator is denoted as $\hat{\mathbf{\Psi}}_{\mathit{LS}} = [\hat{\mathbf{\Phi}}_{\mathit{LS}}, \operatorname{vec}(\hat{\mathbf{C}}_{\mathit{LS}})]$.

\begin{theorem}[Asymptotic Properties of Unconstrained Estimators] \label{thm:unconstrained_asymptotic}
Let $\{\mathbf{X}_t\}_{t\in\mathbb{Z}}$ be a \textup{Z-MINAR(1)} process satisfying $\rho(\mathbf{A})\rho(\mathbf{B}) < 1$. Under Assumptions \ref{assum:parameter_space} to \ref{assum:moment}, the unconstrained estimator $\hat{\mathbf{\Psi}}_{\mathit{LS}}$ is strongly consistent and asymptotically normal, satisfying:
\[
\sqrt{T} \operatorname{vec}(\hat{\mathbf{\Psi}}_{\mathit{LS}} - \mathbf{\Psi}) \xrightarrow{d} \mathcal{N}\left(\mathbf{0}, \ \mathbf{\Sigma}_{\mathit{uncon}} \right), \quad \text{as } T \to \infty,
\]
where $\mathbf{\Sigma}_{\mathit{uncon}} = \mathbf{H}^{-1} \mathbf{\Omega} \mathbf{H}^{-1}$, $\mathbf{H} = \mathbb{E}[\mathbf{z}_{t-1} \mathbf{z}_{t-1}^\top] \otimes \mathbf{I}_{mn \times mn}$, and $\mathbf{\Omega} = \mathbb{E}\left[ (\mathbf{z}_{t-1} \otimes \mathbf{I}_{mn \times mn}) \mathbf{\Sigma}_{U, t-1} (\mathbf{z}_{t-1}^\top \otimes \mathbf{I}_{mn \times mn}) \right]$. Here $\mathbf{z}_{t-1} = (\operatorname{vec}(\mathbf{X}_{t-1})^\top, 1)^\top$ and $\mathbf{\Sigma}_{U, t-1} = \operatorname{Cov}(\operatorname{vec}(\mathbf{U}_t) \mid \mathcal{F}_{t-1})$ is the conditional covariance matrix of the residual $\mathbf{U}_t$.
\end{theorem}

Following the framework established by \citet{xu2024high}, the derivation of the structural estimators relies on the continuous mapping theorem applied to the Singular Value Decomposition (SVD) projection. To explicitly formalize the asymptotic covariance matrix, we define $\boldsymbol{\alpha} = \operatorname{vec}(\mathbf{A})$, $\boldsymbol{\beta} = \operatorname{vec}(\mathbf{B})$, and the normalized vector $\boldsymbol{\beta}^{(1)} = \boldsymbol{\beta} / \|\mathbf{B}\|_F$. Furthermore, let $\mathbf{P} \in \mathbb{R}^{m^2 n^2 \times m^2 n^2}$ be the permutation matrix associated with the rearrangement operator $\mathcal{G}$ such that $\operatorname{vec}(\mathcal{G}(\mathbf{\Phi})) = \mathbf{P} \operatorname{vec}(\mathbf{\Phi})$.

\begin{theorem}[Asymptotic Properties of PROJ Estimators] \label{thm:z_minar_asymptotic}
Under the conditions of Theorem \ref{thm:unconstrained_asymptotic} and the identifiability constraint $\|\mathbf{A}\|_F = 1$, the PROJ estimators $\hat{\mathbf{A}}_{\mathit{PROJ}}$, $\hat{\mathbf{B}}_{\mathit{PROJ}}$, and $\hat{\mathbf{C}}_{\mathit{PROJ}}$ are strongly consistent and jointly satisfy the following asymptotic normal distribution:
\[
\sqrt{T}\begin{pmatrix} 
\operatorname{vec}(\hat{\mathbf{A}}_{\mathit{PROJ}}-\mathbf{A}) \\ 
\operatorname{vec}(\hat{\mathbf{B}}_{\mathit{PROJ}}-\mathbf{B}) \\ 
\operatorname{vec}(\hat{\mathbf{C}}_{\mathit{PROJ}}-\mathbf{C}) 
\end{pmatrix} \xrightarrow{d} \mathcal{N}(\mathbf{0}, \ \mathbf{\Xi}), \quad \text{as } T \to \infty,
\]
where the asymptotic covariance matrix is explicitly given by $\mathbf{\Xi} = \mathbf{J}^* \mathbf{\Sigma}_{\mathit{uncon}} {\mathbf{J}^*}^\top$. The augmented Jacobian matrix $\mathbf{J}^*$ takes the block-diagonal form:
\[
\mathbf{J}^* = \begin{pmatrix} \mathbf{J}_{SVD}\mathbf{P} & \mathbf{0} \\ \mathbf{0} & \mathbf{I}_{mn \times mn} \end{pmatrix}, \quad \text{with} \quad \mathbf{J}_{SVD} = \begin{pmatrix} \|\mathbf{B}\|_F^{-1} \left[ \boldsymbol{\beta}^{(1)\top} \otimes (\mathbf{I}_{m^2 \times m^2} - \boldsymbol{\alpha} \boldsymbol{\alpha}^\top) \right] \\ \mathbf{I}_{n^2 \times n^2} \otimes \boldsymbol{\alpha}^\top \end{pmatrix}.
\]
\end{theorem}

\begin{remark}
While Theorem \ref{thm:z_minar_asymptotic} assumes the true parameter matrices strictly reside within the interior of the positive parameter space ($\mathbb{R}_{>0}$) to establish asymptotic normality theoretically without boundary complications, it is crucial to note that the projection-based estimation method practically accommodates boundary estimates. In finite empirical samples, such as our subsequent regional crime data application, estimated parameters can exactly hit the zero lower bound, accurately reflecting the absence of specific structural spillover effects without invalidating the overall estimation framework.
\end{remark}

\section{Numerical Experiments and Empirical Analysis}
\label{sec:numerical_experiments}

This section evaluates the practical utility of the proposed \textup{Z-MINAR(1)} model through both numerical experiments and a real-world application. First, we use Monte Carlo simulations to assess the finite-sample performance of the projection-based conditional least squares (PROJ) estimator, comparing it with the standard least squares (LS) estimator applied to the fully vectorized \textup{Z-VINAR(1)} framework. The simulation evaluation is conducted across different temporal lengths $T$, matrix dimensions $(m, n)$, and innovation structures. Subsequently, we demonstrate the empirical efficacy of the \textup{Z-MINAR(1)} model by applying it to regional crime count data to uncover dynamic spatiotemporal dependence structures.   

\subsection{Simulation Studies}
\label{subsec:simulation}

We generate the matrix-valued time series $\mathbf{X}_t$ from the \textup{Z-MINAR(1)} model. To ensure stationarity and identifiability, the true coefficient matrices $\mathbf{A}$ and $\mathbf{B}$ are generated randomly under the constraints $\rho(\mathbf{A})\rho(\mathbf{B}) < 1$ and $\|\mathbf{A}\|_F = 1$. These matrices are kept constant across all replications for a given dimension. An initial 500 observations are simulated and discarded as a burn-in period to eliminate transient effects.

We consider three covariance structures for the extended Poisson innovations $\mathbf{E}_t \sim \text{E-Po}(p_{i,j}, \lambda_{i,j})$:

\begin{itemize}
    \item Setting I (Homogeneous Independence): The covariance matrix $\operatorname{Cov}(\operatorname{vec}(\mathbf{E}_t)) = \mathbf{\Sigma}$ is diagonal. Parameters are set to $\lambda_{i,j} = 5$ and $p_{i,j} = 0.5$ for all elements.
    
    \item Setting II (Heterogeneous Independence): $\mathbf{\Sigma}$ is diagonal but heterogeneous. The parameters are drawn from uniform distributions: $\lambda_{i,j} \sim U(2,10)$ and $p_{i,j} \sim U(0.3,0.7)$.
    
    \item Setting III (Dense Cross-Sectional Dependence): To test robustness against structural misspecification, we introduce a diagonal matrix of absolute standard normal variables $\mathbf{\Lambda}$, and a random orthonormal matrix $\mathbf{Q}$. We use a dense covariance structure $\mathbf{\Sigma} = \mathbf{Q}\mathbf{\Lambda}\mathbf{Q}^\top$. We set $p_{i,j} \sim U(0.3,0.7)$.
\end{itemize}

We examine sample sizes $T \in \{100, 200, 500, 1000\}$ and matrix dimensions $(m,n) \in \{(3,2), (6,4), (12,8)\}$. For each setting, we run $n_{rep} = 30$ independent replications. Estimation errors are measured by logarithmic squared Frobenius norms: $\log\|\hat{\mathbf{B}}_{\mathit{PROJ}} \otimes \hat{\mathbf{A}}_{\mathit{PROJ}} - \mathbf{B} \otimes \mathbf{A}\|_F^2$ for PROJ, $\log\|\hat{\mathbf{\Phi}}_{\mathit{LS}} - \mathbf{B} \otimes \mathbf{A}\|_F^2$ for LS, and $\log(\|\hat{\mathbf{C}}_{\mathit{PROJ}} - \mathbf{C}\|_F^2 / \|\mathbf{C}\|_F^2)$ for the intercept.

\begin{figure}[H]
    \centering
    \includegraphics[width=0.89\textwidth]{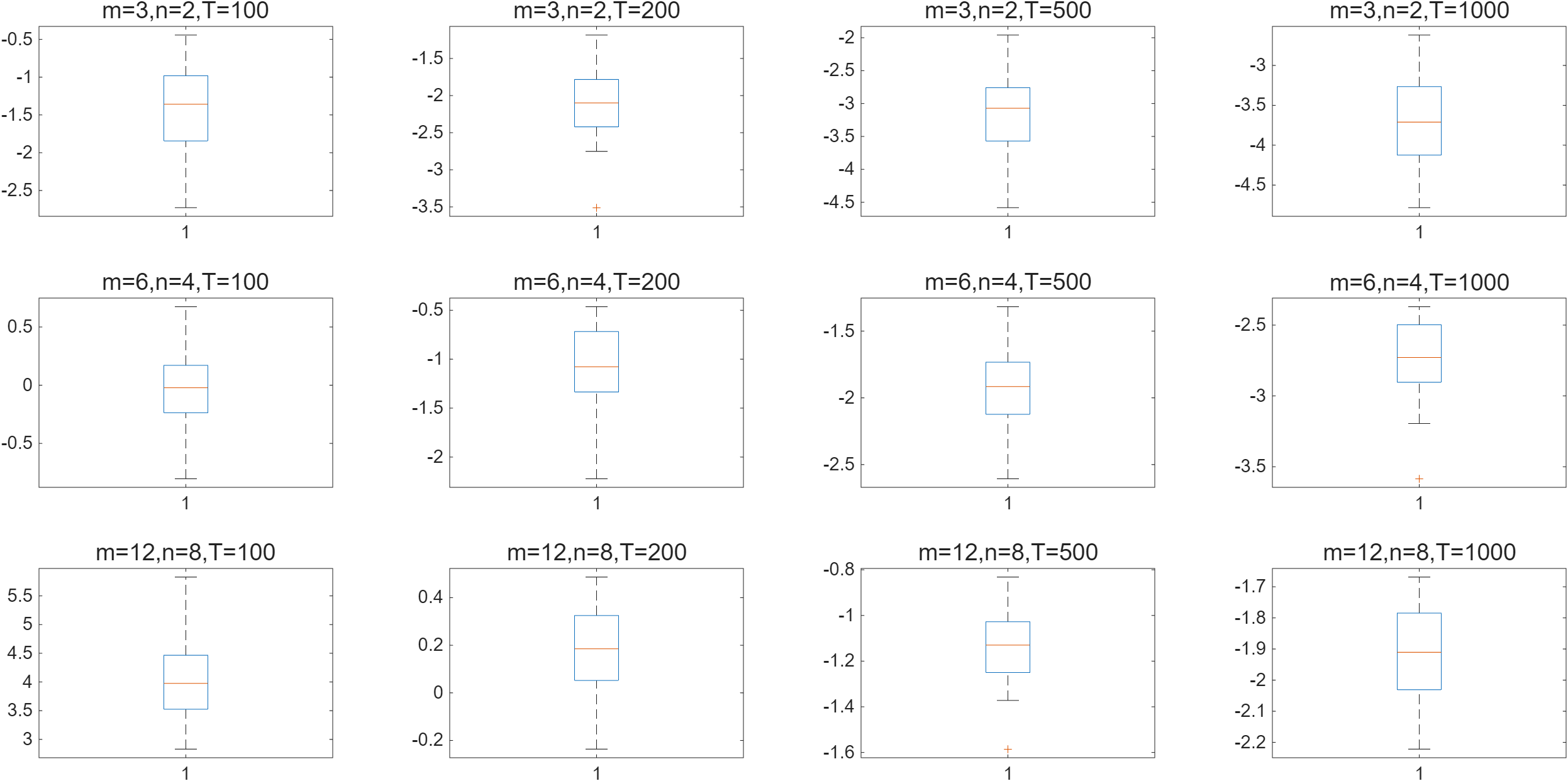}
    \caption{Box plots of the relative estimation errors for the constant matrix $\mathbf{C}$ under Setting III across varying sample sizes $T$ and matrix dimensions $(m,n)$.}
    \label{fig:box_C}
\end{figure}

The results are consistent across all three settings. Using Setting III as an example, Figure \ref{fig:box_C} demonstrates that for a fixed dimension, the estimation error of the constant matrix $\mathbf{C}$ systematically decreases as the temporal sample size $T$ increases. Conversely, for a fixed $T$, the estimation error inherently increases with the matrix dimension due to the larger number of intercept parameters that need to be simultaneously estimated. Specifically, an examination of the box plots reveals that the interquartile range (IQR) of the relative errors narrows noticeably as $T$ grows from 100 to 1000. This shrinking variance indicates that the PROJ estimator becomes increasingly stable and consistent in large samples. Furthermore, while scaling up the matrix dimension from $(3,2)$ to $(12,8)$ inevitably elevates the overall error baseline, the median error still exhibits a strict, monotonic downward trend with respect to $T$. This behavior confirms that the unconstrained estimation of the intercept matrix $\mathbf{C}$ remains asymptotically reliable and statistically sound, even when faced with relatively high-dimensional scenarios and dense cross-sectional dependence in the innovations.

\begin{figure}[H]
    \centering
    \includegraphics[width=0.89\textwidth]{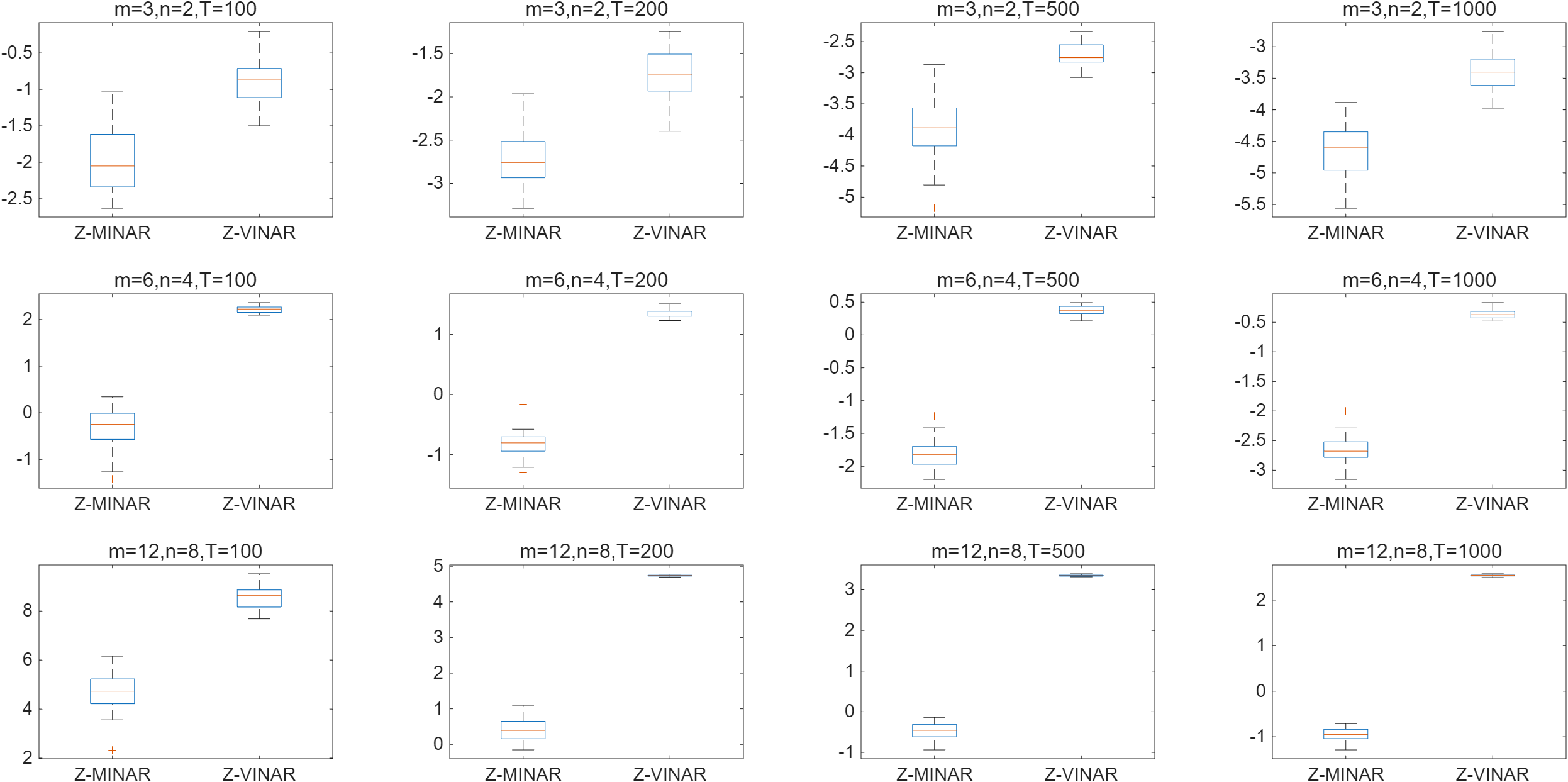}
    \caption{Comparison of the autoregressive coefficient estimation errors between the \textup{Z-MINAR(1)} PROJ estimator and the \textup{Z-VINAR(1)} LS estimator under Setting III.}
    \label{fig:box_coef}
\end{figure}

Figure \ref{fig:box_coef} plots the estimation errors for the composite autoregressive coefficient matrix $\mathbf{B} \otimes \mathbf{A}$. While both estimators demonstrate asymptotic improvement with larger sample sizes $T$, the PROJ estimator significantly and consistently outperforms the baseline LS estimator, particularly as the matrix dimensions scale up. The unconstrained LS estimator suffers severely from the curse of dimensionality, leading to substantially larger biases and error variances. This performance gap becomes strikingly evident in the $(12,8)$ dimensional case, where the LS estimator exhibits extremely high estimation errors, rendering it practically ineffective and unstable for finite samples. The severe degradation of the unconstrained method underscores the absolute necessity of structural regularization in high-dimensional discrete time series modeling. In contrast, the PROJ estimator inherently benefits from the Kronecker product constraint. By projecting the unstructured parameter matrix onto the Kronecker manifold via Singular Value Decomposition, the \textup{Z-MINAR} model drastically reduces the degrees of freedom required for estimation. This structural awareness effectively captures the intrinsic row-column interaction dynamics, thereby achieving a significantly tighter error bound and highly robust convergence, even under the heavily misspecified innovation covariance structure of Setting III.

To further examine element-wise performance, we calculate the empirical Bias and Root Mean Square Error (RMSE) for three representative parameters: $\hat{a}_{1,1}$, $\hat{b}_{2,1}$, and $\hat{c}_{1,1}$. We fix the dimension at $(m,n) = (3,2)$ and use $n_{rep} = 1000$ replications.

\begin{table}[H]
    \centering
    \caption{Empirical Bias and RMSE of selected parameter estimates across Settings I, II, and III for dimension $(m,n)=(3,2)$.}
    \label{tab:bias_rmse}
    \setlength{\tabcolsep}{3.5mm}      
    \renewcommand{\arraystretch}{1.2} 
    \begin{tabular}{cc rrc rrc rr}
        \toprule
        \multirow{2}{*}{Setting} & \multirow{2}{*}{$T$} & \multicolumn{2}{c}{$\hat{a}_{1,1}$} & & \multicolumn{2}{c}{$\hat{b}_{2,1}$} & & \multicolumn{2}{c}{$\hat{c}_{1,1}$} \\
        \cline{3-4} \cline{6-7} \cline{9-10}
        & & Bias & RMSE & & Bias & RMSE & & Bias & RMSE \\
        \midrule
        \multirow{3}{*}{I}   
        & 200  & -0.0168 & 0.0722 & & 0.0052 & 0.0658 & & 0.0494 & 0.4244 \\
        & 500  & -0.0078 & 0.0409 & & -0.0005 & 0.0413 & & 0.0040 & 0.2567 \\
        & 1000 & -0.0004 & 0.0289 & & 0.0029 & 0.0294 & & 0.0244 & 0.1893 \\
        \midrule
        \multirow{3}{*}{II}  
        & 200  & -0.0119 & 0.0693 & & 0.0111 & 0.0721 & & -0.0344 & 0.4706 \\
        & 500  & -0.0094 & 0.0391 & & 0.0062 & 0.0387 & & 0.0276 & 0.3003 \\
        & 1000 & -0.0056 & 0.0296 & & 0.0041 & 0.0266 & & 0.0203 & 0.1998 \\
        \midrule
        \multirow{3}{*}{III} 
        & 200  & -0.0183 & 0.0694 & & -0.0062 & 0.0782 & & -0.0324 & 0.5205 \\
        & 500  & -0.0091 & 0.0414 & & 0.0002 & 0.0459 & & 0.0617 & 0.3385 \\
        & 1000 & -0.0038 & 0.0267 & & 0.0009 & 0.0302 & & 0.0065 & 0.2256 \\
        \bottomrule
    \end{tabular}
\end{table}

As shown in Table \ref{tab:bias_rmse}, both Bias and RMSE decay toward zero as $T$ increases across all settings, aligning with our theoretical results. Notably, under the dense cross-sectional dependence in Setting III, the PROJ estimator remains largely unbiased. This indicates that the SVD projection effectively isolates the row and column structures despite heavy misspecification in the innovation covariance.

\begin{figure}[H]
    \centering
    \includegraphics[width=0.89\textwidth]{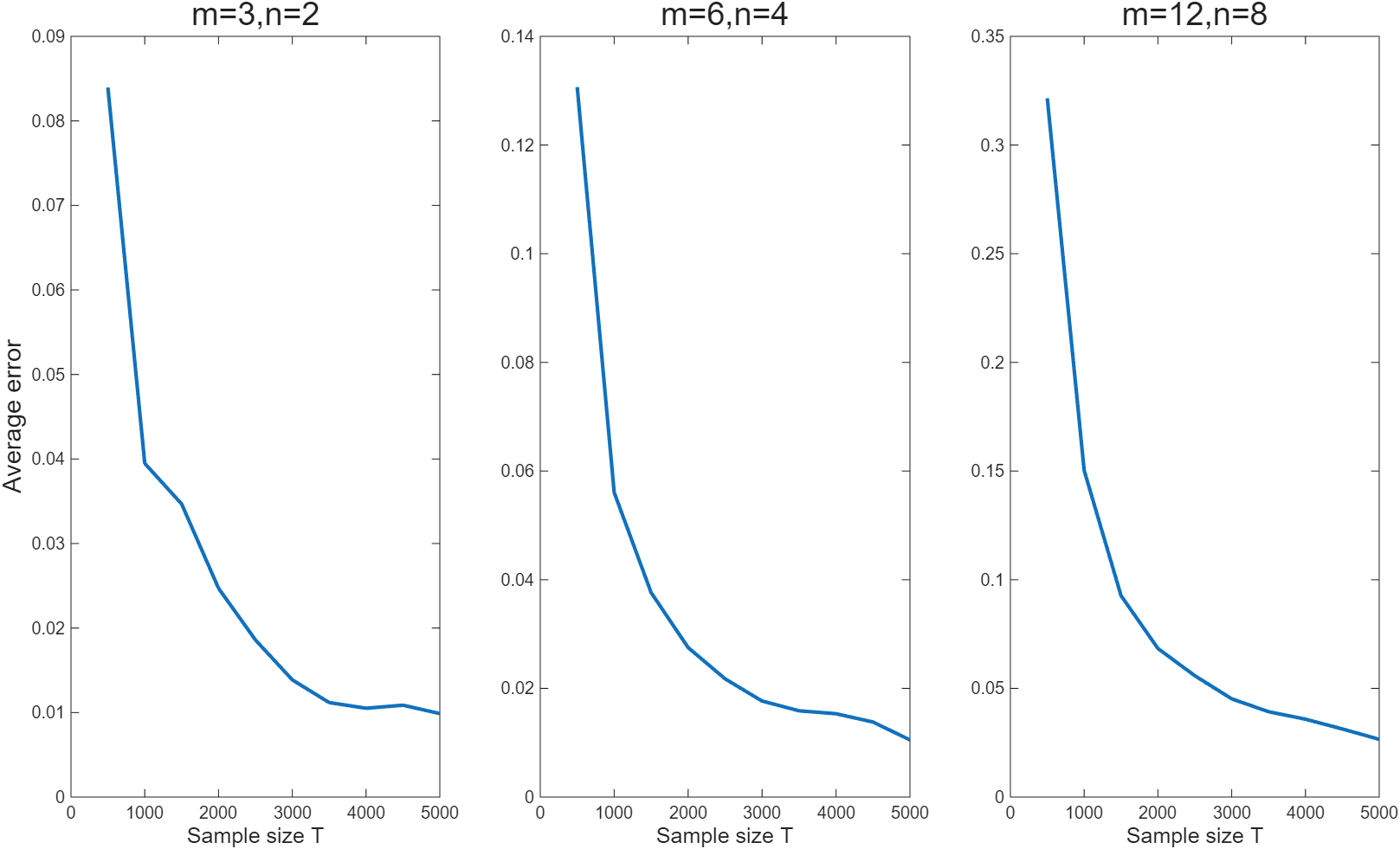}
    \caption{Average estimation error for the constant matrix $\mathbf{C}$ under Setting III as sample size $T$ increases.}
    \label{fig:asy_C}
\end{figure}

\begin{figure}[H]
    \centering
    \includegraphics[width=0.89\textwidth]{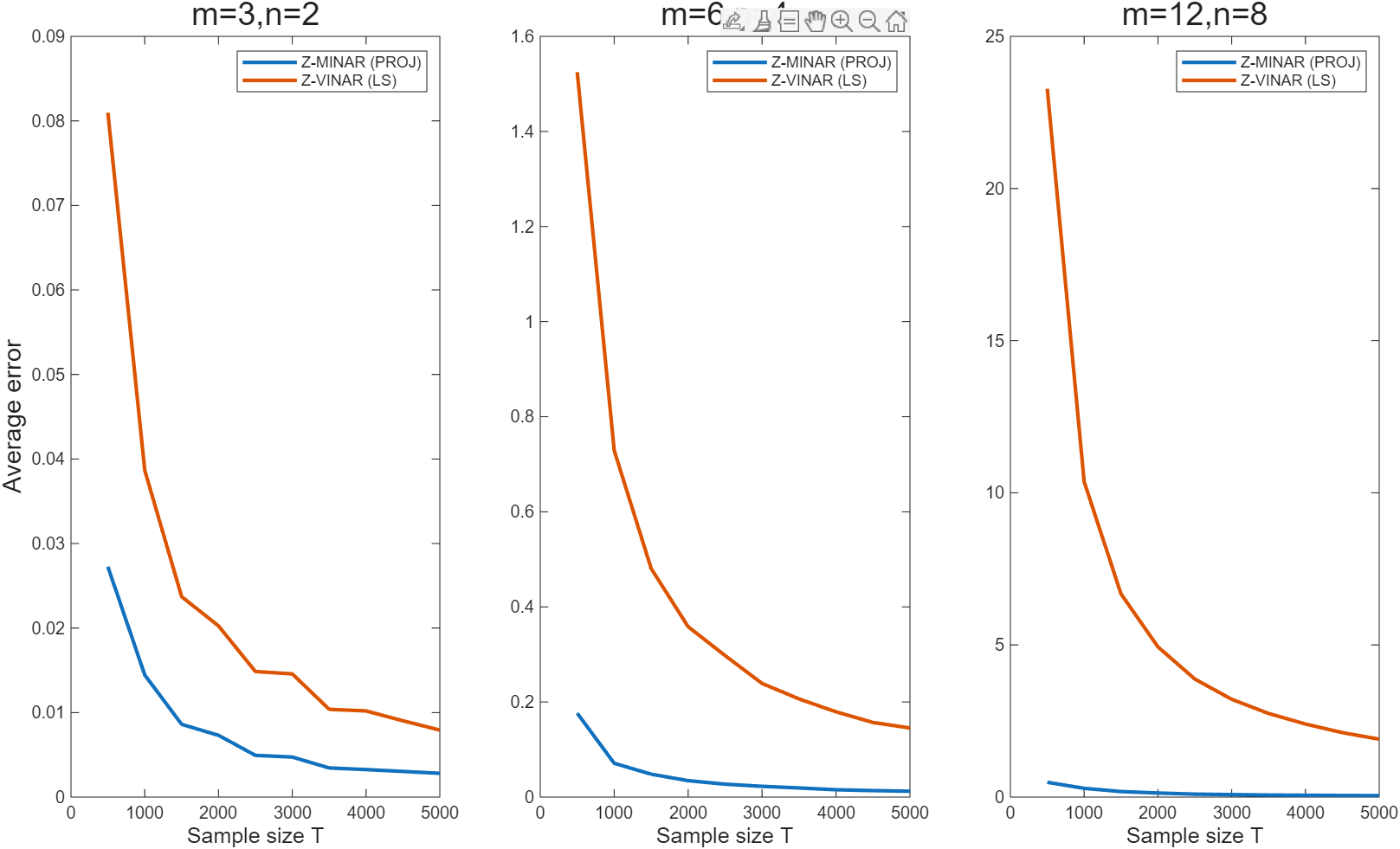}
    \caption{Comparison of asymptotic efficiencies for the coefficient matrix $\mathbf{B} \otimes \mathbf{A}$ between PROJ and LS estimators under Setting III.}
    \label{fig:asy_coef}
\end{figure}

To assess asymptotic efficiency, we increase the sample size up to $T = 5000$. Figures \ref{fig:asy_C} and \ref{fig:asy_coef} demonstrate that the PROJ estimator's error decays at a noticeably faster rate than the unstructured vector autoregressive baseline, particularly as the matrix dimensions scale up.

Finally, we compute the empirical coverage probabilities of the 95\% confidence intervals to verify Theorem \ref{thm:z_minar_asymptotic}. Based on 1000 replications for $(m,n)=(3,2)$, we evaluate two parameter vectors: $(\operatorname{vec}(\hat{\mathbf{A}}_{\mathit{PROJ}})^\top, \operatorname{vec}(\hat{\mathbf{B}}_{\mathit{PROJ}})^\top)^\top$ and $\operatorname{vec}(\hat{\mathbf{B}}_{\mathit{PROJ}} \otimes \hat{\mathbf{A}}_{\mathit{PROJ}})$. 

\begin{table}[H]
    \centering
    \caption{Empirical coverage probabilities of 95\% confidence intervals.}
    \label{tab:coverage}
    \setlength{\tabcolsep}{6mm}      
    \renewcommand{\arraystretch}{1.3} 
    \begin{tabular}{ccccc}
        \toprule
        & Setting & I & II & III \\
        \midrule
        \multirow{3}{*}{$(\operatorname{vec}(\hat{\mathbf{A}}_{\mathit{PROJ}})^\top, \operatorname{vec}(\hat{\mathbf{B}}_{\mathit{PROJ}})^\top)^\top$}
        & $T=200$  & 0.8758 & 0.8006 & 0.8120 \\
        & $T=500$  & 0.9245 & 0.9203 & 0.8982 \\
        & $T=1000$ & 0.9406 & 0.9104 & 0.9165 \\
        \midrule
        \multirow{3}{*}{$\operatorname{vec}(\hat{\mathbf{B}}_{\mathit{PROJ}} \otimes \hat{\mathbf{A}}_{\mathit{PROJ}})$}
        & $T=200$  & 0.8681 & 0.7909 & 0.7959 \\
        & $T=500$  & 0.8806 & 0.9114 & 0.8639 \\
        & $T=1000$ & 0.9220 & 0.9112 & 0.9056 \\
        \bottomrule
    \end{tabular}
\end{table}

\FloatBarrier

\subsection{Empirical Analysis: Regional Crime Count Variations}
\label{subsec:empirical_analysis}

According to criminological theory, variations in crime rates are not isolated events; they exhibit complex spatiotemporal interactions driven by geographical proximity, socio-economic shifts, and law enforcement interventions. To capture these dynamic interactions, we examine crime data from four distinct police districts in Chicago (District 11, District 14, District 15, and District 25). We specifically focus on the semi-monthly counts of three major crime types: CRIMINAL DAMAGE, BATTERY, and THEFT, recorded between January 1, 2001, and December 31, 2022.

\begin{figure}[H]
    \centering
    \includegraphics[width=0.87\textwidth]{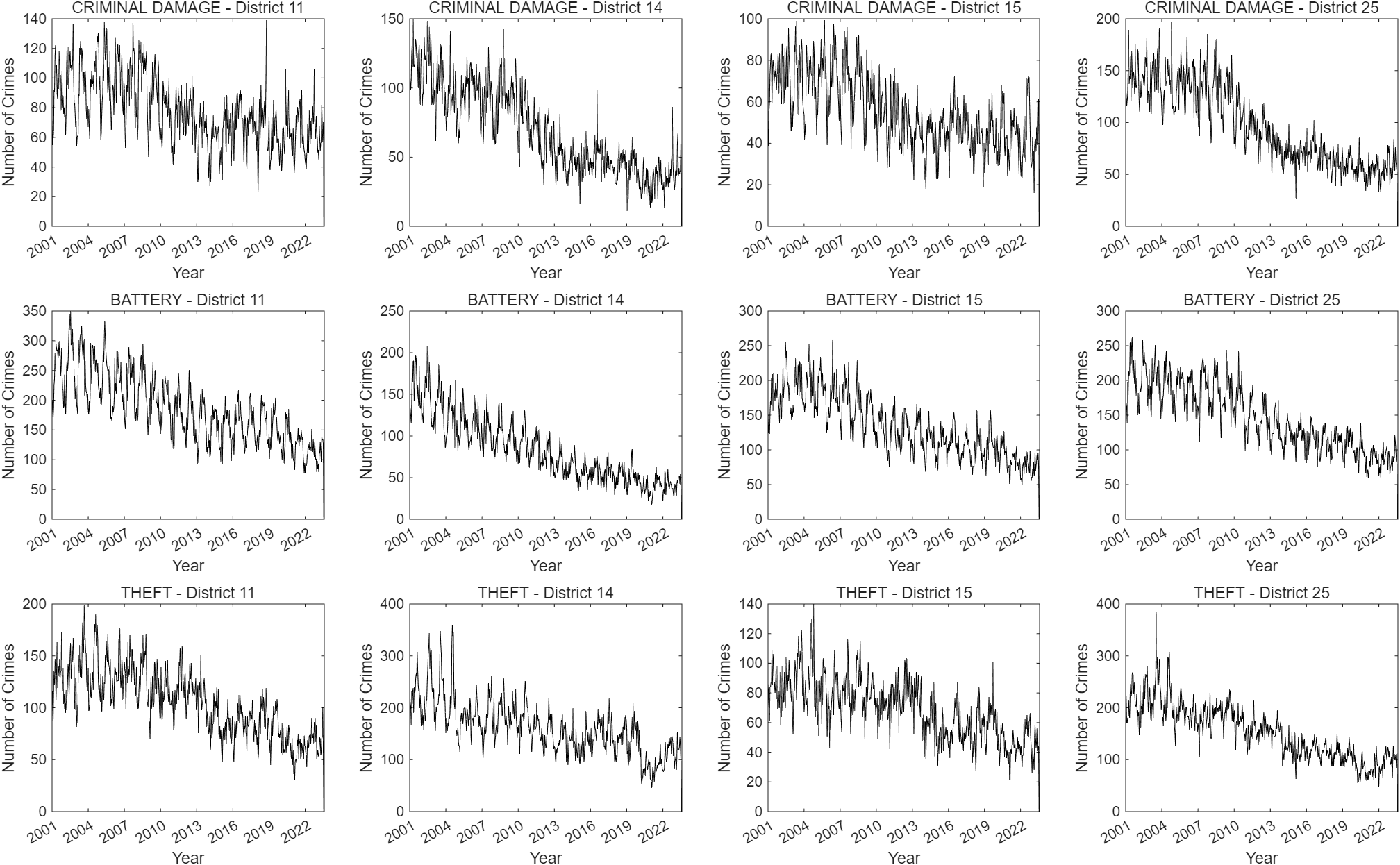}
    \caption{Time series of three types of crime (CRIMINAL DAMAGE, BATTERY, and THEFT) in the four selected districts of Chicago.}
    \label{fig:crime_timeseries}
\end{figure}

As visually evidenced in Figure \ref{fig:crime_timeseries}, the raw semi-monthly counts exhibit significant temporal volatility and varying baseline frequencies across different crime types and spatial locations. For instance, THEFT generally occurs at a substantially higher volume compared to CRIMINAL DAMAGE and BATTERY across all districts. Moreover, the sequences display high-frequency semi-monthly fluctuations and occasional spikes, underscoring the stochastic and episodic nature of local crime occurrences. Such inherent volatility and cross-sectional heterogeneity further motivate our methodological approach: rather than modeling absolute levels, analyzing the period-to-period dynamic variations allows for a more robust extraction of the underlying spatiotemporal dependence structures.

\begin{figure}[H]
    \centering
    \includegraphics[width=0.87\textwidth]{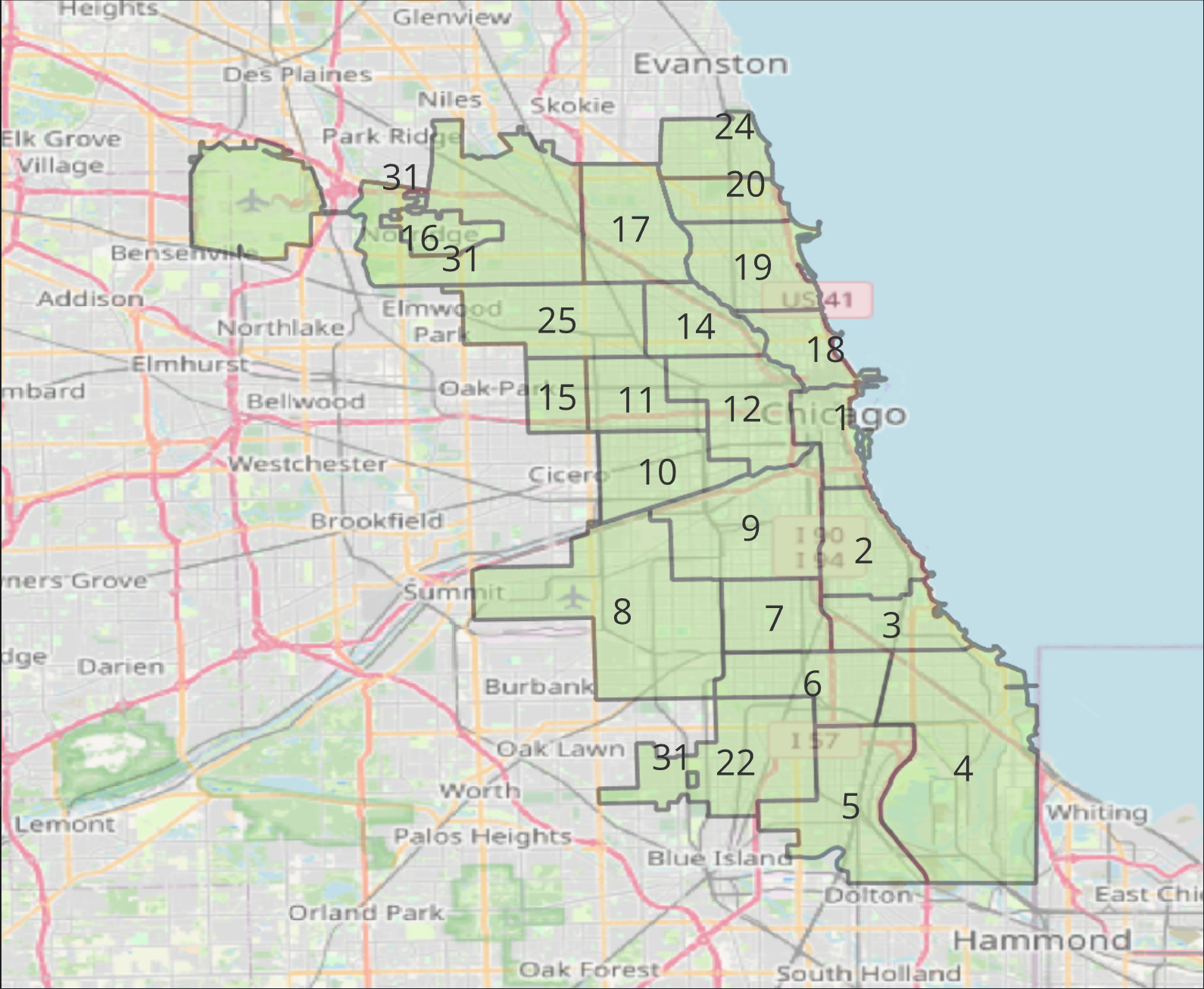}
    \caption{Geographical locations of the selected police districts (District 11, District 14, District 15, and District 25) in Chicago.}
    \label{fig:chicago_map}
\end{figure}

Figure \ref{fig:chicago_map} illustrates the geographical distribution of these four selected police districts, highlighting their relative spatial proximity which serves as the physical basis for the cross-sectional dependence analyzed in our model.

While traditional count time series models focus on absolute non-negative occurrences, evaluating the \textit{dynamic changes} and \textit{fluctuations} in crime---such as period-to-period surges or drops---is critical for assessing the immediate impact of police patrols or short-term environmental factors. Therefore, we transform the original non-negative dataset, denoted as $\mathbf{Y}_t$, by taking the first temporal difference of the semi-monthly counts. Following the notation introduced in Section \ref{sec:z-minar1}, the differenced series is defined as $\mathbf{X}_t = \nabla \mathbf{Y}_t = \mathbf{Y}_t - \mathbf{Y}_{t-1}$. This differencing process fundamentally maps the data onto the entire integer domain ($\mathbb{Z}$), yielding both positive increments and negative reductions. At any given time $t$, the differenced observation $\mathbf{X}_t$ forms a $3 \times 4$ matrix with entries in $\mathbb{Z}$. The total length of the differenced time series is $T = 541$. We allocate the first 432 observations as the training set to estimate the model parameters, and retain the remaining 109 observations as the test set to evaluate out-of-sample forecasting performance.

Because the differenced matrix data contains negative integers, traditional models such as \textup{MGINAR} and \textup{MINAR} are mathematically rendered inapplicable due to their strict reliance on the classical non-negative binomial thinning operator. To provide meaningful benchmarks, we adapt the traditional models using the signed thinning operator and compare the proposed \textup{Z-MINAR(1)} model against the following modified vector/matrix frameworks:
\begin{enumerate}[(i)]
    \item \textup{Z-VINAR}\_row(1): Fits three independent Vector $\mathbb{Z}$-valued \textup{INAR(1)} models to each row of the matrix series, modeling the variations of the three crime types separately across districts.
    \item \textup{Z-VINAR}\_column(1): Fits four independent Vector $\mathbb{Z}$-valued \textup{INAR(1)} models to each column, modeling the crime variations within each district separately.
    \item iZ-INAR(1): Fits 12 completely independent univariate $\mathbb{Z}$-valued \textup{INAR(1)} models for each specific crime type in each district.
    \item \textup{Z-VINAR(1)} (LS): Fits the full unstructured multivariate $\mathbb{Z}$-valued \textup{INAR(1)} model to the vectorized data $\operatorname{vec}(\mathbf{X}_t)$ using standard least squares. This model serves as a highly parameterized benchmark, acting as a direct vector extension of the univariate framework proposed by \citet{kachour2023new}, employing an unstructured parameter matrix $\mathbf{\Phi}$ without the Kronecker constraints.
    \item \textup{Z-MINAR(1)} (PROJ): The proposed model, fitting the matrix data directly using the signed matrix thinning operator and estimated via the projection-based conditional least squares method.
    \item \textup{MINAR(1)} (Non-differenced): Fits the standard non-negative \textup{MINAR(1)} model directly to the original count data $\mathbf{Y}_t$. To ensure a fair comparison, the resulting predictions $\hat{\mathbf{Y}}_t$ are explicitly converted to differenced predictions via $\hat{\mathbf{X}}_t = \hat{\mathbf{Y}}_t - \mathbf{Y}_{t-1}$.
\end{enumerate}

The training and prediction performances are rigorously evaluated using three standard error metrics (we omit mean absolute percentage error as the true values in $\mathbb{Z}$ frequently include zeros):
\begin{align*}
e_1 &= \sum_{t=1}^{T} \sqrt{\sum_{i=1}^{m}\sum_{j=1}^{n} (x_{i,j,t} - \hat{x}_{i,j,t})^2}, \\
e_2 &= \sqrt{\frac{1}{mnT} \sum_{t=1}^{T}\sum_{i=1}^{m}\sum_{j=1}^{n} (x_{i,j,t} - \hat{x}_{i,j,t})^2}, \\
e_3 &= \frac{1}{mnT} \sum_{t=1}^{T}\sum_{i=1}^{m}\sum_{j=1}^{n} |x_{i,j,t} - \hat{x}_{i,j,t}|.
\end{align*}

Table \ref{tab:empirical_insample} summarizes the in-sample fitting results on the training set. The proposed \textup{Z-MINAR(1)} model demonstrates a highly efficient structural representation, drastically reducing the parameter count to 37 by leveraging the Kronecker product approximation. Despite this substantial parameter reduction, \textup{Z-MINAR(1)} achieves highly competitive in-sample fitting metrics, efficiently capturing the complex underlying dynamics of the differenced crime data. In comparison, while the fully unstructured \textup{Z-VINAR(1)} yields slightly lower numerical errors due to its massive parameter space (156 parameters), it is highly susceptible to overfitting. The other baseline models, such as \textup{iZ-INAR(1)} and the partially structured \textup{Z-VINAR}\_row and \textup{Z-VINAR}\_column, offer varying degrees of trade-offs but lack the elegant bidirectional (spatial-typological) interaction structure explicitly captured by \textup{Z-MINAR(1)}.

  \begin{table}[H]
      \centering
      \caption{Comparison of in-sample fitting effects of the models on differenced crime data.}
      \label{tab:empirical_insample}
      \setlength{\tabcolsep}{4mm}      
      \renewcommand{\arraystretch}{1.2} 
      \begin{tabular}{l cccc}
          \toprule
          Model & $e_1$ & $e_2$ & $e_3$ & No. of Parameters \\
          \midrule
          \textup{Z-VINAR}\_row(1) & 27870.1529 & 19.6353 & 15.0064 & 60 \\
          \textup{Z-VINAR}\_column(1) & 27926.2806 & 19.7043 & 15.0173 & 48 \\
          iZ-INAR(1) & 28079.1223 & 19.8163 & 15.1355 & 24 \\
          \textup{Z-VINAR(1)} (LS) & \textbf{27539.0715} & \textbf{19.4253} & \textbf{14.8301} & 156 \\
          \textup{Z-MINAR(1)} (PROJ) & 27936.0039 & 19.6932 & 15.0526 & 37 \\
          \textup{MINAR(1)} (Non-differenced) & 41608.4451 & 28.8556 & 23.0248 & 37 \\
          \bottomrule
      \end{tabular}
  \end{table}

The true robustness of the models is revealed in their out-of-sample forecasting performance, as presented in Table \ref{tab:empirical_outsample}. The \textup{Z-MINAR(1)} model stands out by demonstrating exceptional predictive capabilities on the differenced data ($\nabla \mathbf{Y}_t$). By efficiently modeling the intrinsic row-column interaction dynamics between crime spikes and drops, \textup{Z-MINAR(1)} successfully extracts meaningful spatial and typological spillovers without falling into the trap of overfitting. In contrast, the heavily parameterized \textup{Z-VINAR(1)} suffers from the "curse of dimensionality," leading to degraded forecasting accuracy. While the non-differenced \textup{MINAR(1)} yields lower numerical errors by acting as a strong smoothing baseline on highly autocorrelated raw counts, it fails to model the actual short-term fluctuations. Other differenced baselines, such as the disconnected \textup{iZ-INAR(1)} and the partially structured \textup{Z-VINAR}\_row and \textup{Z-VINAR}\_column models, show competitive numerical results but lack the comprehensive structural interpretability that \textup{Z-MINAR(1)} provides through its Kronecker product framework. Ultimately, \textup{Z-MINAR(1)} provides the most balanced and interpretable approach for forecasting complex matrix-variate crime fluctuations.

  \begin{table}[H]
      \centering
      \caption{Comparison of the out-of-sample forecasting effects of the models on differenced crime data.}
      \label{tab:empirical_outsample}
      \setlength{\tabcolsep}{4mm}      
      \renewcommand{\arraystretch}{1.2} 
      \begin{tabular}{l cccc}
          \toprule
          Model & $e_1$ & $e_2$ & $e_3$ & No. of Parameters \\
          \midrule
          \textup{Z-VINAR}\_row(1) & 5489.6458 & 16.4085 & 11.8107 & 60 \\
          \textup{Z-VINAR}\_column(1) & 5463.9436 & 16.3951 & 11.7732 & 48 \\
          iZ-INAR(1) & 5465.9271 & 16.4269 & 11.7948 & 24 \\
          \textup{Z-VINAR(1)} (LS) & 5515.6991 & 16.4741 & 11.8563 & 156 \\
          \textup{Z-MINAR(1)} (PROJ) & \textbf{5452.2195} & \textbf{16.3359} & \textbf{11.7640} & 37 \\
          \textup{MINAR(1)} (Non-differenced) & 7888.0101 & 22.4914 & 17.5419 & 37 \\
          \bottomrule
      \end{tabular}
  \end{table}

\begin{figure}[H]
    \centering
    \includegraphics[width=0.8\textwidth]{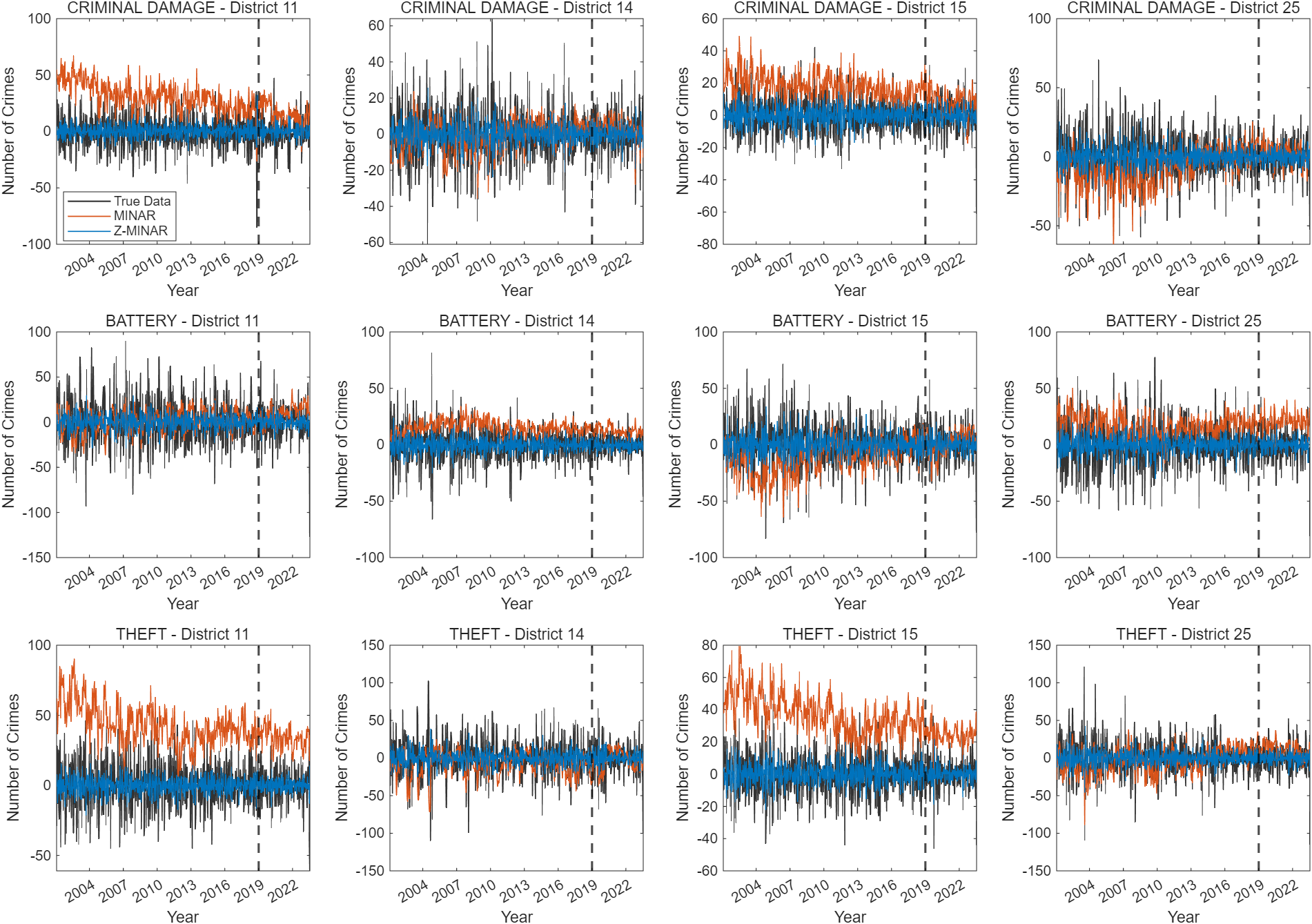}
    \caption{Comparison between the actual fluctuations of differenced crime counts (True Data, black line) and the predicted trajectories of the \textup{Z-MINAR} (blue line) and non-differenced \textup{MINAR} (orange line) models across different crimes and districts. The area to the left of the black dotted line represents the in-sample fitting period, while the right side represents the out-of-sample forecasting period.}
    \label{fig:crime_prediction}
\end{figure}

To provide a more intuitive visual verification of the \textup{Z-MINAR(1)} model's dynamic tracking capability at a micro-temporal scale, we plot the actual differenced series of all three crime types across the four districts. Figure \ref{fig:crime_prediction} illustrates this comprehensive comparison, contrasting the true data against the predicted trajectories from both our proposed \textup{Z-MINAR(1)} model and the unconstrained \textup{Z-VINAR} baseline.

From the time series trajectories in Figure \ref{fig:crime_prediction}, several key observations emerge. First, the non-differenced \textup{MINAR} model (orange line) produces highly flattened predictions near zero. Because it models the non-stationary raw counts directly, its predictions strongly anchor to the immediate past absolute levels ($\mathbf{Y}_t \approx \mathbf{Y}_{t-1}$), effectively treating most semi-monthly variations as unpredictable noise. As a result, it completely fails to capture any meaningful dynamic volatility or sharp fluctuations, despite yielding a numerically lower MSE by acting as a conservative smoothing baseline. In stark contrast, the proposed \textup{Z-MINAR(1)} model (blue line) actively traces the fluctuation patterns of the differenced data. During extreme peaks (sharp surges or sudden drops), \textup{Z-MINAR(1)} executes rapid directional adjustments and successfully anticipates the scale and timing of local extreme fluctuations across the 3x4 grid. This visual evidence strongly corroborates the practical necessity of the differencing step and underscores \textup{Z-MINAR}'s unique structural capability in uncovering dynamic spatial and typological spillovers.

To gain deeper criminological insights, Table \ref{tab:empirical_matrices} displays the parameter matrices $\hat{\mathbf{A}}_{\mathit{PROJ}}$, $\hat{\mathbf{B}}_{\mathit{PROJ}}$, and $\hat{\mathbf{C}}_{\mathit{PROJ}}$ estimated by the \textup{Z-MINAR(1)} model. Due to the inherent scale ambiguity of the Kronecker product ($\mathbf{B} \otimes \mathbf{A} = (c\mathbf{B}) \otimes (\mathbf{A}/c)$), the absolute magnitudes of the individual matrices $\mathbf{A}$ and $\mathbf{B}$ are not uniquely identified. To strictly ensure parameter identifiability as established in Theorem 4.2, we enforce the normalization constraint $\|\hat{\mathbf{A}}_{\mathit{PROJ}}\|_F = 1$. Consequently, our empirical analysis fundamentally focuses on the \textit{relative spillover intensity}---the proportional distribution of structural influence across different crime types and spatial districts---rather than interpreting their absolute numerical scales.

The row interaction matrix $\hat{\mathbf{A}}_{\mathit{PROJ}}$ reveals an interesting cross-crime dynamic: while the crime variations are heavily self-dependent (as evidenced by the dominant diagonal values around 0.56 to 0.60), there is very little cascading spillover effect among these specific crime types. Specifically, initial surges in CRIMINAL DAMAGE have a very slight leading effect on BATTERY (0.0002), while THEFT remains almost entirely independent of the others. This highlights that these particular property and violent crimes tend to fluctuate largely independently of one another.

The column interaction matrix $\hat{\mathbf{B}}_{\mathit{PROJ}}$ highlights the spatial transmission of crime dynamics among the districts. Specifically, District 11 and District 14 exhibit a strong bidirectional spillover influence, with District 11 influencing District 14 (0.1206) and District 14 influencing District 11 (0.0971). Additionally, District 15 strongly influences District 14 (0.1428), while District 25 also exerts influence on District 11 (0.0688) and 14 (0.0621). This dense and asymmetric spatial interaction feature reveals a highly interconnected network of crime volatility spillovers in these regions, providing a crucial empirical basis for formulating coordinated, cross-district preventive police deployments. 

Finally, the relatively small magnitudes in the drift matrix $\hat{\mathbf{C}}_{\mathit{PROJ}}$ confirm that, after differencing, the overall baseline trend of the crime counts remains relatively stable around zero.

  \begin{table}[H]
      \centering
      \caption{PROJ Estimates of parameter matrices $\hat{\mathbf{A}}_{\mathit{PROJ}}$, $\hat{\mathbf{B}}_{\mathit{PROJ}}$, and $\hat{\mathbf{C}}_{\mathit{PROJ}}$ for the \textup{Z-MINAR(1)} model.$^{a}$}
      \label{tab:empirical_matrices}
      \setlength{\tabcolsep}{1.5mm}      
      \renewcommand{\arraystretch}{1.2} 
      \begin{tabular}{ll cccc}
          \toprule
          \multicolumn{2}{l}{Parameter matrix} & \multicolumn{4}{c}{Estimates} \\
          \midrule
          & & CRIMINAL DAMAGE & BATTERY & THEFT & \\
          \multirow{3}{*}{$\hat{\mathbf{A}}_{\mathit{PROJ}}$} 
          & CRIMINAL DAMAGE  & 0.5990 & 0.0000 & 0.0000 & \\
          & BATTERY          & 0.0002 & 0.5683 & 0.0000 & \\
          & THEFT            & 0.0000 & 0.0000 & 0.5642 & \\
          \midrule
          & & District 11 & District 14 & District 15 & District 25 \\
          \multirow{4}{*}{$\hat{\mathbf{B}}_{\mathit{PROJ}}$} 
          & District 11 & 0.0000 & 0.0971 & 0.0736 & 0.0688 \\
          & District 14 & 0.1206 & 0.0000 & 0.1428 & 0.0621 \\
          & District 15 & 0.0987 & 0.0801 & 0.0000 & 0.0025 \\
          & District 25 & 0.1270 & 0.1109 & 0.0211 & 0.0000 \\
          \midrule
          & & District 11 & District 14 & District 15 & District 25 \\
          \multirow{3}{*}{$\hat{\mathbf{C}}_{\mathit{PROJ}}$} 
          & CRIMINAL DAMAGE  &  0.0648 & -0.1783 & -0.0374 & -0.1433 \\
          & BATTERY          & -0.1556 & -0.2967 & -0.0942 & -0.2053 \\
          & THEFT            & -0.0704 & -0.1514 & -0.0405 & -0.3595 \\
          \bottomrule
      \end{tabular}
      \vspace{1ex}
      \begin{flushleft}
      \footnotesize $^{a}$ Due to scale ambiguity, $\hat{\mathbf{A}}_{\mathit{PROJ}}$ is normalized to have a Frobenius norm of 1.
      \end{flushleft}
  \end{table}

While the fixed out-of-sample testing in the previous section provides an intuitive evaluation of predictive accuracy, its static nature might be sensitive to the specific time period chosen. To rigorously verify the statistical robustness of the proposed \textup{Z-MINAR(1)} model's predictive superiority, we design a one-step-ahead rolling window forecasting experiment coupled with the Diebold-Mariano (DM) test. 

We maintain a constant training window size of $w = 432$ periods. For each out-of-sample time step $t$ (from period 433 to 541), the model parameters are recursively re-estimated using the historical data from $t-w$ to $t-1$. The updated models are then used to forecast the matrix variations for period $t$. This procedure yields a dynamic sequence of 109 out-of-sample prediction error matrices for both the structured \textup{Z-MINAR(1)} (PROJ) model and the fully parameterized \textup{Z-VINAR(1)} (LS) baseline.

To assess whether the accuracy improvements are statistically significant, we employ the DM test under the null hypothesis of equal predictive accuracy. We calculate the loss differentials utilizing both the Squared Error (SE) loss and the Absolute Error (AE) loss aggregated over the matrix elements. The results are summarized in Table \ref{tab:dm_test}.

\begin{table}[H]
\centering
\caption{Diebold-Mariano (DM) Test Results for Predictive Accuracy (\textup{Z-VINAR(1)} LS versus \textup{Z-MINAR(1)} PROJ)}
\label{tab:dm_test}
\setlength{\tabcolsep}{6mm}
\renewcommand{\arraystretch}{1.3}
\begin{tabular}{lcc}
\toprule
 & Squared Error (SE) Loss & Absolute Error (AE) Loss \\
\midrule
DM Statistic & 1.4169 & 1.0053 \\
$p$-value & 0.1565 & 0.3148 \\
\bottomrule
\end{tabular}
\end{table}

As demonstrated in Table \ref{tab:dm_test}, the DM test statistics for both squared and absolute error losses are positive, indicating that the proposed PROJ estimator yields a smaller average loss than the benchmark LS estimator. However, the $p$-values (0.1565 and 0.3148) suggest that this difference is not statistically significant at the conventional 5\% level. This result implies that while dynamically enforcing the Kronecker product constraint via projection successfully mitigates the curse of dimensionality and achieves a numerically lower forecasting error (as seen in Table \ref{tab:empirical_outsample}), the highly stochastic nature of the semi-monthly spatiotemporal count series makes the statistical distinction between the two estimators marginal. Nonetheless, the \textup{Z-MINAR(1)} model maintains a more parsimonious structure (37 parameters vs. 156) without sacrificing predictive accuracy.

\FloatBarrier

\section{Conclusion}

This paper aims to break through the theoretical and practical bottlenecks faced by existing integer-valued time series models when dealing with complex, high-dimensional matrix data containing negative values. To this end, we innovatively propose a matrix autoregressive model defined on the full integer domain ($\mathbb{Z}$)---the \textup{Z-MINAR(1)} model. By pioneering a signed matrix thinning operator and integrating extended Poisson innovations, this research successfully shatters the structural limitations of traditional \textup{MINAR} models, which are strictly confined to non-negative integers. The proposed framework not only flexibly accommodates differenced fluctuation data prevalent in the real world but also precisely captures the intrinsic row and column cross-correlations among variables through its Kronecker product structure.

In terms of parameter estimation and statistical inference, this paper develops a projection-based conditional least squares (PROJ) estimation method. Compared to traditional unconstrained vectorized \textup{INAR} models, this approach effectively overcomes the ``curse of dimensionality'' inherent in high-dimensional spatiotemporal data by extracting the dominant singular value for manifold projection, thereby significantly reducing the computational burden and parameter redundancy. More importantly, we establish a rigorous theoretical foundation for the model by formally proving its white noise representation, stationarity, causality, and the asymptotic normality of the projection estimators, ensuring the validity of statistical inference under large samples.

Extensive numerical experiments and empirical analyses demonstrate the outstanding practical value of the proposed model. Monte Carlo simulation results indicate that under various settings of heterogeneity and complex cross-sectional dependence, the PROJ estimation method exhibits superior finite-sample accuracy, strong robustness, and asymptotic efficiency that far exceeds benchmark models. Furthermore, in the empirical application to regional crime count fluctuations, the \textup{Z-MINAR(1)} model, thanks to its parsimonious structure, successfully avoids overfitting. It comprehensively outperforms competing models in out-of-sample forecasting and intuitively uncovers the complex spatiotemporal dynamic spillover networks between different crime types (row effects) and geographical regions (column effects).

Despite its theoretical and empirical advantages, the \textup{Z-MINAR(1)} model still has some limitations that provide broad prospects for future methodological extension. First, the current model is limited to first-order autocorrelation. Future research should generalize the model to higher-order lag structures, developing a \textup{Z-MINAR}($p$) model to capture long-memory temporal dynamics, which naturally introduces the challenge of determining the appropriate lag order $p$ for matrix-valued integer data. Second, the current framework does not incorporate external information. Integrating macroscopic exogenous covariates into the model would further enhance its capability in analyzing large-scale spatiotemporal data driven by multiple external factors.

\newpage
\bibliographystyle{elsarticle-num-names}
\bibliography{1}
\end{document}